\documentclass[aps,twocolumn]{revtex4-1}

\usepackage{bm, amsmath,amssymb,amsfonts,slashed,array,graphicx,soul}
\usepackage[vcentermath,noautoscale]{youngtab}
\usepackage{mathrsfs}
\usepackage{xcolor}
\usepackage{hyperref}
\usepackage{cancel}
\usepackage[normalem]{ulem}
\DeclareMathAlphabet{\mathpzc}{OT1}{pzc}{m}{it}

\Yboxdim{8pt}

\newcommand{\nc}{\newcommand}

\nc{\beq}{\begin{equation}} \nc{\eeq}{\end{equation}}
\nc{\beqa}{\begin{eqnarray}} \nc{\eeqa}{\end{eqnarray}}
\nc{\bea}{\begin{eqnarray}} \nc{\eea}{\end{eqnarray}}
\nc{\barray}{\begin{eqnarray}} \nc{\earray}{\end{eqnarray}}
\nc{\barrayn}{\begin{eqnarray*}} \nc{\earrayn}{\end{eqnarray*}}
\nc{\ra}{\rightarrow}
\nc{\lsim}{\begin{array}{c}\,\sim\vspace{-21pt}\\< \end{array}}
\nc{\gsim}{\begin{array}{c}\sim\vspace{-21pt}\\> \end{array}}
\nc{\Tr}{{\rm Tr}} \nc{\slsh}{\slash\hspace*{-0.22cm}}

\def\be{\begin{equation}}
\def\ee{\end{equation}}
\def\bea{\begin{eqnarray}}
\def\eea{\end{eqnarray}}
\def\bit{\begin{itemize}}
\def\eit{\end{itemize}}

\nc{\infinity}{\infty} \nc{\mc}{\mathcal} \nc{\M}{\mathcal{M}}

\def\lsim{\mathrel{\rlap{\lower4pt\hbox{\hskip1pt$\sim$}}
    \raise1pt\hbox{$<$}}}
\def\gsim{\mathrel{\rlap{\lower4pt\hbox{\hskip1pt$\sim$}}
    \raise1pt\hbox{$>$}}}

\def\to{\rightarrow}

\begin{document}

\vbox{\baselineskip14pt
} {~~~~~~~~~~~~~~~~~~~~~~~~~~~~~~~~~~~~
~~~~~~~~~~~~~~~~~~~~~~~~~~~~~~~~~~~
 \footnotesize{MCTP-15-20}}

\title{Dark Matter Annihilation Decay at The LHC}
\author{Yuhsin Tsai$^{1}$, Lian-Tao Wang$^{2,3}$, and Yue Zhao$^{4}$}
\affiliation{$^1$Maryland Center for Fundamental Physics, Department of Physics, University of Maryland, College Park, MD 20742}
\affiliation{$^2$Department of Physics, The University of Chicago,
Chicago, IL 60637} \affiliation{$^3$Enrico Fermi Institute and Kavli
Institute for Cosmological Physics, The University of Chicago,
Chicago, IL 60637} \affiliation{$^4$Michigan Center for Theoretical
Physics, University of Michigan, Ann Arbor, MI 48109}
\begin{abstract}
Collider experiments provide an opportunity to shed light on dark
matter (DM) self-interactions. In this work, we study the
possibility of generating DM bound states -- the Darkonium -- at the
LHC and discuss how the annihilation decay of the Darkonium produces
force carriers. We focus on two popular scenarios that contain large
DM self-couplings: the Higgsinos in the $\lambda$-SUSY model, and
self-interacting DM (SIDM) framework. After forming bound states,
the DM particles annihilate into force mediators, which decay into
the standard model particles either through a prompt or displaced
process. This generates interesting signals for the heavy resonance
search. We calculate the production rate of bound states and study
the projected future constraints from the existing heavy resonance
searches.
\end{abstract}
\maketitle

\section{Introduction}
The existence of dark matter (DM) has been proven by many
astrophysical observations. However, all the supportive evidence so
far comes from gravitational interactions between DM and standard
model (SM) particles, and there is no direct and unambiguous
evidence of other types of DM  interactions yet. Many efforts, such
as the direct and indirect detection experiments, have been devoted
to search for the non-gravitational DM-SM couplings. With no clear
discoveries so far, it is important to look for other types of
experiments that can provide a complimentary search.

Collider experiments serve this purpose well. Among many
advantages of looking for DM particles at colliders
\cite{Birkedal:2004xn,Bai:2010hh,Goodman:2010ku,An:2015pva}, one
unique feature of these high energy experiments is that they provide
a chance to study the mediators of DM interactions. Many papers have
discussed looking for the mediator particles between the dark and SM
sectors (see \cite{An:2012ue,Bai:2013iqa,Liu:2013gba} for example).
In this work, we instead study how the production of DM bound states
at the Large Hadron Collider (LHC) can help to shed light on the DM
self-interaction, which is mediated by a force carrier that couples
strongly to the DM particle but weakly to the SM sector.

In this paper, we focus on two well-motivated scenarios that may
naturally have strong DM self-interactions: the self-interacting
dark matter models (SIDM) and $\lambda$-SUSY. DM self-interaction
can impact the structures of DM halos \cite{Spergel:1999mh}. Several
astrophysical observations show potential deviations from theoretical
predictions if only gravitational interaction of DM particles are
included \cite{DWcore,TBTF1,TBTF2,MWslope}. Moreover, detailed
simulations show that some anomalies can be resolved by having
self-interacting DM (SIDM) with a scattering cross section
$\sigma/m_{\chi}\sim 0.1-10\,\text{cm}^2/\text{g}$ between DM
particles \cite{simulation1,simulation2,simulation3}. Such a large
cross section implies a strong DM self-interaction, which can come
from the mediation of a light force carrier \cite{Tulin:2013teo}. 
If the self interaction is strong enough, DM bound states can be formed at colliders from the pair production of DM particles. 

Likewise, the Higgs boson has been discovered at the LHC with mass
around 125 GeV \cite{Aad:2012tfa,Chatrchyan:2012xdj}. The mass of
the Higgs boson is too heavy to be naturally explained in the
minimal supersymmetric SM (MSSM). With a variation to the
superpotential, the $\lambda$-SUSY scenario introduces a large
F-term to the Higgs potential which helps to raise the Higgs mass
\cite{Barbieri:2006bg,Hall:2011aa}. The large value of $\lambda$
implies a sizable attractive Yukawa coupling among Higgsinos and
Singlino, mediated by the singlet scalar and Higgs boson. If the
lightest neutralino, which serves as a good candidate for DM, is
dominantly Higgsino and Singlino, the light force scalar mediator
can also help to form a bound state when neutralinos or charged
Higgsinos are pair produced.

A high energy collider provides special access to DM
self-interactions. Once DM particles are produced near the
threshold, there is a chance for them to form a bound state, the
Darkonium, due to a strong self-interaction \footnote{The name
Darkonium in the DM context was first used in \cite{Laha:2015yoa}
for a stable DM bound state in an asymmetric DM model. Weakly coupled DM bound states have been studied in \cite{Pospelov:2008jd, MarchRussell:2008tu, Kaplan:2009de, Braaten:2013tza,Wise:2014jva}.}. The
particle and anti-particle in the bound state can easily find each
other and annihilate into light mediators or SM particles. The
collider signatures of forming a dark matter bound state is very
different from the traditional DM search at colliders
\cite{Beltran:2010ww,Fox:2011fx,Fox:2011pm,Rajaraman:2011wf,Carpenter:2012rg,Lin:2013sca}.
Instead of looking for missing energy and its recoiled objects, one
can look for the resonance of the bound state. This greatly reduces
the SM background and helps to extract information about the dark
sector, such as reconstructing the mass of DM particles.
Furthermore, the self-interaction force mediators can have a small
coupling to the SM sector and negligible direct production rates at
colliders. The bound state production, however, can generate the
force mediators from the bound state annihilation decay, which
provides a plausible way to study the mediators \footnote{Also see
\cite{Gupta:2015lfa,Buschmann:2015awa,An:2015pva} for another
example of producing mediators from the DM production.}.

Here we note that the collider production of DM bound state by the
weakly interacted massive particles (WIMP) has been discussed in
\cite{Shepherd:2009sa}. Different from their approach, we also study
the possible bound state production in $\lambda$-SUSY, which is a
well-motivated scenario. Further, we study the SIDM scenario with
the input of preferred values of DM self-interaction strength, based
on simulations. In Sec.\ref{sec:ReviewBoundState}, we review the
basics on calculations of bound state production rate at a hadron
collider. In Sec.\ref{sec:lamdaSUSY}, we focus on the $\lambda$-SUSY
scenario. Bound states are formed by neutralino/chargino. The
neutralino/chargino production is through W/Z bosons and the
singlet/Higgs bosons behave as the force mediators to form bound
states. The annihilation decay products can be the mediator or W/Z
bosons. We perform a simple PDF rescaling on existing similar
searches at the LHC in order to get a rough estimation on the reach
limit at higher energy and larger luminosity.  In
Sec.\ref{sec:sidm}, we study the bound state from the SIDM model.
The production of DM particles is calculated using effective
operators, similar to the assumption in MET searches. However we
emphasize that the constraints on the mediation scale of effective
operator in MET searches can sometimes be much lower than the
typical energy scale of the collision. Thus there are concerns of
self-consistency using the language of low energy effective theory.
This problem is relieved in our scenario because the energy scale of
our process is fixed to be the bound state mass. Finally, we
summarize in Sec.\ref{sec:summary}.

\section{Bound state production}\label{sec:ReviewBoundState}
We follow the technique in \cite{Peskin:1995ev,Kats:2009bv,Kahawala:2011pc} to
calculate the bound state production at the LHC. When the
DM particles in a bound state are produced near the threshold, we can
write the binding energy and wave function in terms of the DM mass $m$ and self-coupling
$\alpha_{\lambda}$. For the $s$-wave bound state,
we have
\begin{eqnarray}\label{eq:boundrate}
 E_b&=&\frac{\alpha_\lambda^2\,m}{4},\quad a_0^{-1}=\frac{1}{2}\,\alpha_{\lambda}\,m,\\\left|\psi(0)\right|^2&=& \frac{1}{\pi a_0^3}= \frac{\alpha_{\lambda}^{3}\,m^{3}}{8\pi},\quad\mathcal{M}_b=\sqrt{\frac{1}{m}}\,\psi(0)\,\mathcal{M}_0,\nonumber
 \end{eqnarray}
where $E_b$ is the binding energy, $\psi(0)$ is the wavefunction of
DM particles at zero separation, $\mathcal{M}_0$ is the amplitude of
producing two free DM particles, and $\mathcal{M}_b$ is the matrix
element for generating a bound state. The self-coupling
$\alpha_{\lambda}$ is obtained at the energy scale of the inverse
Bohr radius $a_0^{-1}$. In the $\lambda$-SUSY discussion, we define $\lambda$ to be the Higgsino-singlet Yukawa coupling at the Higgsino mass. Since there is no running of this particular Yukawa coupling below the mass scale, it is the $\alpha_{\lambda}$ to use for the bound state calculation. Similarly, we define the size of the SIDM coupling
at the mediator mass, and there is no sizable running of the
DM-mediator coupling down to the binding energy. We then do not
include the running of the self-couplings in this work.

Eq.~(\ref{eq:boundrate}) gives a non-relativistic approximation of
the DM particles in the bound state. The expression is valid when
the constituent particles in the bound state have speed
$v\sim\alpha_{\lambda}<1$. Further we need the light mediator
wavelength to be longer than the Bohr radius, which requires
$m_{\text{med}}<\alpha_{\lambda}m/2$. The $p$-wave has $\psi(0)=0$,
and the amplitude $\mathcal{M}_b$ depends on the derivative of
$\psi(x)$ with respect to the radial coordinate. This gives a
relative suppression
$|\psi'(0)|^2/|\psi(0)|^2\sim\alpha_{\lambda}^2$ for the bound state
production and decay compared to the s-wave state. We then do not
discuss the $p$-wave or higher angular-excited states in this work.

To obtain the bound state production rate, we first calculate the
amplitude of having free partons $\chi$ scattering into the SM
quarks $\chi\bar{\chi}\to q\bar{q}$. According to the total angular
momentum of the bound state, we use corresponding wavefunction of
each $\chi$ in spinor space in the non-relativistic limit while
keeping the quark wavefunction in the relativistic form. Summing the
non-vanishing combinations of the spin polarization, the decay rate,
$\Gamma_{{\bf B}\to q\bar{q}}$, can be calculated through
$\mathcal{M}_b$. Averaging over the bound state polarization, one
obtains the bound state production cross section as
\cite{Kats:2009bv} 
\begin{equation}\label{eq:BScross}
\sigma(q\bar{q}\to{\bf B})=\zeta(3)\frac{4\pi^2(2J+1)}{9M^3}\,\mathcal{L}_{q\bar{q}}(M^2)\,\Gamma_{{\bf B}\to q\bar{q}},
\end{equation}
where $M\simeq2m$ is the mass of the DM bound state.
Here $\zeta(3)$ comes from summing the modes of radial excitations.
In principle, the sum should stop once the Bohr radius of radial
excitation states is longer than the Compton wavelength of the
mediator. But it only causes a difference of $O(1)$. $J$ is the
total angular momentum of the bound state. For the production at the
LHC, one needs to include the integral of PDF in
$\mathcal{L}_{q\bar{q}}$ with the center of mass energy set to the
mass of the bound state.
\section{Higgsino bound state in $\lambda$-SUSY}\label{sec:lamdaSUSY}
\subsection{Parameters in $\lambda$-SUSY}
In $\lambda$-SUSY, a large value of $\lambda$ helps to increase the
Higgs mass to 125 GeV in a natural way \cite{Hall:2011aa}. The
$\lambda S H_u H_d$ term in the superpotential induces the Yukawa
coupling between Higgsinos/Singlino and the singlet scalar or Higgs
boson, i.e.
\begin{eqnarray}
  \label{eq:Yukawa}
\mathcal{L}&\supset& \lambda \tilde{H}_u \tilde{H}_d s+\lambda
\tilde{H}_u H_d \tilde S+\lambda H_u \tilde{H}_d \tilde S + h.c.,
\end{eqnarray}
where $\tilde{H}$ and $\tilde{S}$ are Higgsinos and Singlino. $s$ is
the scalar component of singlet $S$. Given that $\lambda$ can be
very large in $\lambda$-SUSY, $s$ can mediate a strong attractive
force between higgsinos. Once the Higgsinos are pair produced at the
LHC, such a strong attractive force may induce a Higgsino bound
state. The annihilation decay of this bound state may provide us a
powerful handle on the search for Higgsinos.

Here we study the parameter space in $\lambda$-SUSY, in which the
neutralino bound state is important. Given the fact that $s$ does
not couple to Wino and Bino directly, we would like to focus on the
scenario where the lightest neutralinos are mainly Higgsino and
Singlino. Following the convention in \cite{Ellwanger:2009dp}, the
general NMSSM super potential is written as
\begin{eqnarray}
 \label{eq:Wpot}
W=\lambda S H_u H_d +\xi_F S+\frac{1}{2}\mu' S^2+\frac{\kappa}{3}S^3.
\end{eqnarray}
In the $Z_3$ invariant NMSSM, $\xi_F$ and $\mu'$ are absent. In the
basis $\psi^0=(-i\lambda_1,-i\lambda_2^3,\psi^0_d,\psi^0_u,\psi_S)$,
the neutralino mass matrix is written as ($\mu_{eff}=\lambda s$)
\begin{eqnarray}
M_0= \left(%
\begin{array}{cccccc}
              M_1 & 0 &-\frac{g_1 v_d}{\sqrt{2}} & \frac{g_1 v_u}{\sqrt{2}} & 0&\cr
               & M_2  & \frac{g_2 v_d}{\sqrt{2}} & -\frac{g_2 v_u}{\sqrt{2}} & 0 \cr
               &   &  0 & -\mu_{eff}& -\lambda v_u  \cr
               &   &   & 0  &  -\lambda v_d \cr
               &   &   &   &  2\kappa s +\mu' \cr
\end{array}%
\right).
\end{eqnarray}
If $M_1$ and $M_2$ are large compared to other parameters in the
mass matrix, Wino and Bino are heavy, and their components in the
lightest neutralino are negligible. The singlet scalar can couple to
Higgsinos through a large Yukawa coupling $\lambda$ or to Singlinos
through a large $\kappa$ term.

In principle, it  is possible to form a bound state with an $O(1)$ mixing between Higgsino and Singlino. However, when the mixing is too large, it
reduces the cross section of neutralino production because Singlino
does not couple to the SM such as the $W$ or $Z$ boson. For
simplicity, we assume $(2\kappa s +\mu')\gg\lambda v_{u,d}$ in order
to decouple the Singlino component. We will see in later discussions
that this limit is also helpful to relax the constraints from
electroweak precision tests. Thus in this case, the lightest
neutralino states are mainly Higgsinos with a nearly degenerate
spectrum. This guarantees the decay
$\tilde\chi^0_2\to\tilde\chi^0_1+$ SM to be much slower than the
annihilation decay of the bound state. Since the energy of bound state production is much larger than the splitting
between two neutralinos, we can combine the neutalinos into a Dirac fermion and form an s-wave bound state through the vector coupling of $Z$. The singlet scalar $s$ plays the role of force mediator binding the bound state. If we instead keep only one light neutralino, the $Z$-mediated bound state production of two Majorana fermions is p-wave suppressed.

Furthermore, when $M_2$ is very large, the lightest chargino is mainly a Higgsino, which is
almost degenerate with the lightest neutralinos. A large value of
$\lambda$ also induces a large Yukawa coupling between the singlet and the charged
Higgsinos, which allows the formation of bound states with two
charginos or one chargino and one neutralino.

Finally, let us consider the mass matrix of CP-even scalar
particles. Using the VEVs $v_u$,$v_d$ and $\langle s\rangle$ to
eliminate several soft mass terms in the Lagrangian, the $3\times 3$
mass matrix under the basis of $(H_{d},H_{u},s)$ can be written as
\cite{Allanach:2008qq}
\begin{eqnarray}
 \label{eq:ScalarMatrix}
M_{11}^2&=&g^2 v_d^2 +(\mu_{eff}
B_{eff}+\hat{m}_3^2)\tan\beta\nonumber\\
M_{22}^2&=&g^2 v_u^2 +(\mu_{eff} B_{eff}+\hat{m}_3^2)/\tan\beta\nonumber\\
M_{33}^2&=&\lambda(A_\lambda+\mu')\frac{v_u v_d}{\langle s\rangle}+\kappa \langle s\rangle(A_\kappa+4\kappa \langle s\rangle+3\mu')\nonumber\\&-&(\xi_S+\xi_F\mu')/\langle s\rangle\nonumber\\
M_{12}^2&=&(2\lambda^2-g^2)v_u v_d-\mu_{eff}B_{eff}-\hat{m}_3^2\nonumber\\
M_{13}^2&=&\lambda(2\mu_{eff}v_d-(B_{eff}+\kappa \langle s\rangle+\mu')v_u)\nonumber\\
M_{23}^2&=&\lambda(2\mu_{eff}v_u-(B_{eff}+\kappa \langle s\rangle+\mu')v_d)
\end{eqnarray}
where $g^2=\frac{g_1^2+g^2_2}{2}$, $B_{eff} = A_\lambda+\kappa
\langle s\rangle$, $\hat{m}_3^2 = m_3^2+\lambda\,(\mu'\langle
s\rangle+\xi_F)$, and $m_3$ $A_\lambda$, $A_\kappa$, $\xi_S$ are the
soft SUSY breaking parameters in a general NMSSM. It is quite
involved to do a complete analysis on the possible parameter region.
Here we will only discuss the desired parametrization of this mass
matrix for the bound state production and point out some subtleties
of it. \footnote{We thank very helpful discussions with Bibhushan
Shakya on this point.}

First, let us rotate the mass matrix in Eq.~(\ref{eq:ScalarMatrix})
and study the physics in the basis $(h_v^0,H_v^0,h_s^0)$ defined as
\begin{eqnarray}
  \label{eq:CPevenMatrix}
H_u^0&=&v_u+\frac{1}{\sqrt{2}}(\sin\beta \,h_v^0+\cos\beta \,H_v^0),\nonumber\\
H_d^0&=&v_d+\frac{1}{\sqrt{2}}(\cos\beta \,h_v^0-\sin\beta \,H_v^0),\nonumber\\
s&=&\langle s\rangle+\frac{1}{\sqrt{2}}\,h_s^0.
\end{eqnarray}
Only $h_v^0$ has tree level coupling with $W$ and $Z$ bosons under
this basis. Given the fact that the lightest neutralino/chargino
are dominantly Higgsinos, the most appealing scenario for the bound
state production is the small mixing limit, where the singlet scalar
only has a small mixing with other scalars. This is because a mixing
in the singlet scalar will reduce its Yukawa coupling to Higgsinos.
On the other hand, the mixing between $h_v^0$ and $H_v^0$ is
strongly constrained by the LHC data while the mixing between
$h_v^0$ and $h_s^0$ can be moderate \cite{Farina:2013fsa}.

This forces us to suppress all possible mixings in this mass matrix.
However, when having no sizable mixing from other states, the
tree-level mass of the SM-like Higgs can be approximated as
\begin{eqnarray}
  \label{eq:HiggsMass}
m_{h_v^0}^2\sim \lambda^2 v^2 \sin^2 2\beta+m_Z^2 \cos^2 2\beta,
\end{eqnarray}
and if $\lambda\gsim 2$, the Higgs mass is too large when
$\tan\beta\simeq 1$. If the singlet scalar is heavy, e.g.
$m_{s}\simeq 400$ GeV, only a small mixing between $(h_s^0,\,
h_v^0)$ is required to drive the Higgs mass down, but then the
Higgsino mass needs to be quite large ($m_{\tilde
\chi}>2m_s/\alpha_\lambda$) in order to form the bound state.

There are several ways to get around the above issues while having
light Higgsinos bound state production. One is to consider the
parameter region with $\tan\beta \gsim 5$, so the tree level mass of
the SM-like Higgs is reduced. The model with a large $\tan\beta$ and
$\lambda\sim 2$ can be highly constrained by EW Precision Tests
(EWPT) \cite{Franceschini:2010qz}. There are three types of
loop-contributions to the EWPT: Higgsino/Singlino, stop/sbottom, and
CP-odd/even Higgs bosons. In the $Z_3$ invariant version of NMSSM,
contributions from the Higgsino loop diagrams can be significant at
large $\tan\beta$. However, in a general NMSSM model, a larger
$\mu'$ in Eq. (\ref{eq:Wpot}) can raise the Singlino mass and
suppress the Higgsino-Singlino mixing. This reduces the tension from
EWPT. The contributions from the other two EWPT-violation channels
are more moderate. In the $Z_3$-invariant NMSSM, the stop-sbottom
contribution can be reduced by increasing the heavier charged Higgs
mass $m_{H^\pm}$ \cite{Barbieri:2006bg}. Large soft masses of the
stop and sbottom can also help to loosen the constraints, although a
sizable tuning will be required. A large $m_{H^\pm}$ will also
reduce constraints from the CP-odd/even Higgs bosons, which are
generally weaker in the $Z_3$-invariant NMSSM
\cite{Franceschini:2010qz}.

Besides having a larger $\tan\beta$, the other possibility to obtain
the right Higgs mass and a light singlet scalar is to introduce
another gauge singlet chiral supermultiplet, for example,
$W\supset\lambda SH_uH_d+\mu SS'$. The scalar component $s'$ can be
heavy and play the role of lowering the SM-like Higgs mass to the
observed value, while the singlet scalar $s$ can be light and be the
force mediator to form the Higgsino bound state \footnote{See
\cite{Lu:2013cta} for a similar setup.}.

Finally, we emphasize that in $\lambda$-SUSY, the coupling between
$H_u$, $H_d$ and $S$ is generically sizable. If the singlet mass is
smaller than half the mass of the Higgs, it may induce a large decay
branching ratio of $h\rightarrow ss$, unless properly tuning
parameters to get a small coupling. Thus we require $s$ to be
heavier than 62 GeV. On the other hand, $s$ cannot be too heavy or
else we cannot treat it as a light mediator to form the bound state
of Higgsinos. This induces an upper limit for $s$ mass, i.e. it
should be smaller than the Bohr radius of the bound state
$r_{Bohr}\sim \frac{2}{m_{\tilde h}\alpha_\lambda}$. Thus we assume
\begin{eqnarray}
 \label{eq:MS}
\frac{m_h}{2}< m_s<\frac{m_{\tilde h}\alpha_\lambda}{2}
\end{eqnarray}
when calculating the bound state production.
\subsection{Higgsinonium production}
The bound state production is proportional to the wave function
$\left|\psi(0)\right|^2$, and it is more plausible to discuss the
s-wave ($\ell=0$) state, which has a non-vanishing wave function at
the origin. There are two s-wave bound states to be formed with two
distinguishable fermions: $0^{-}$ and $1^{-}$ in the convention
$J^{P}$ for the spin ($J$) and parity ($P$). The pseudo-scalar
$0^{-}$ can be produced through a gluon fusion of the CP odd scalar
in the complex scalar $S$, and the vector $1^{-}$ can be produced
through the transverse component of the SM gauge bosons. Since the
$0^{-}$ production depends on the details of the scalar spectrum,
and the cross section from the loop-induced gluon fusion process is
relatively small, we focus on the $1^{-}$ production in this work.

\begin{figure}
\center
\includegraphics[width=8cm]{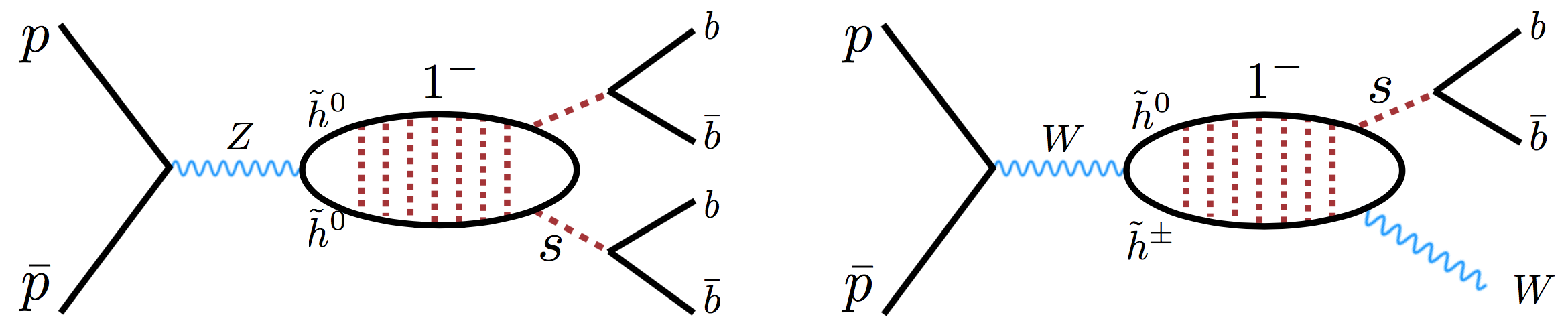}
\caption{Higgsinonium production and decay at the LHC.}\label{fig:diagmHiggs}
\end{figure}

\begin{figure*}
\center
\includegraphics[width=18.5cm]{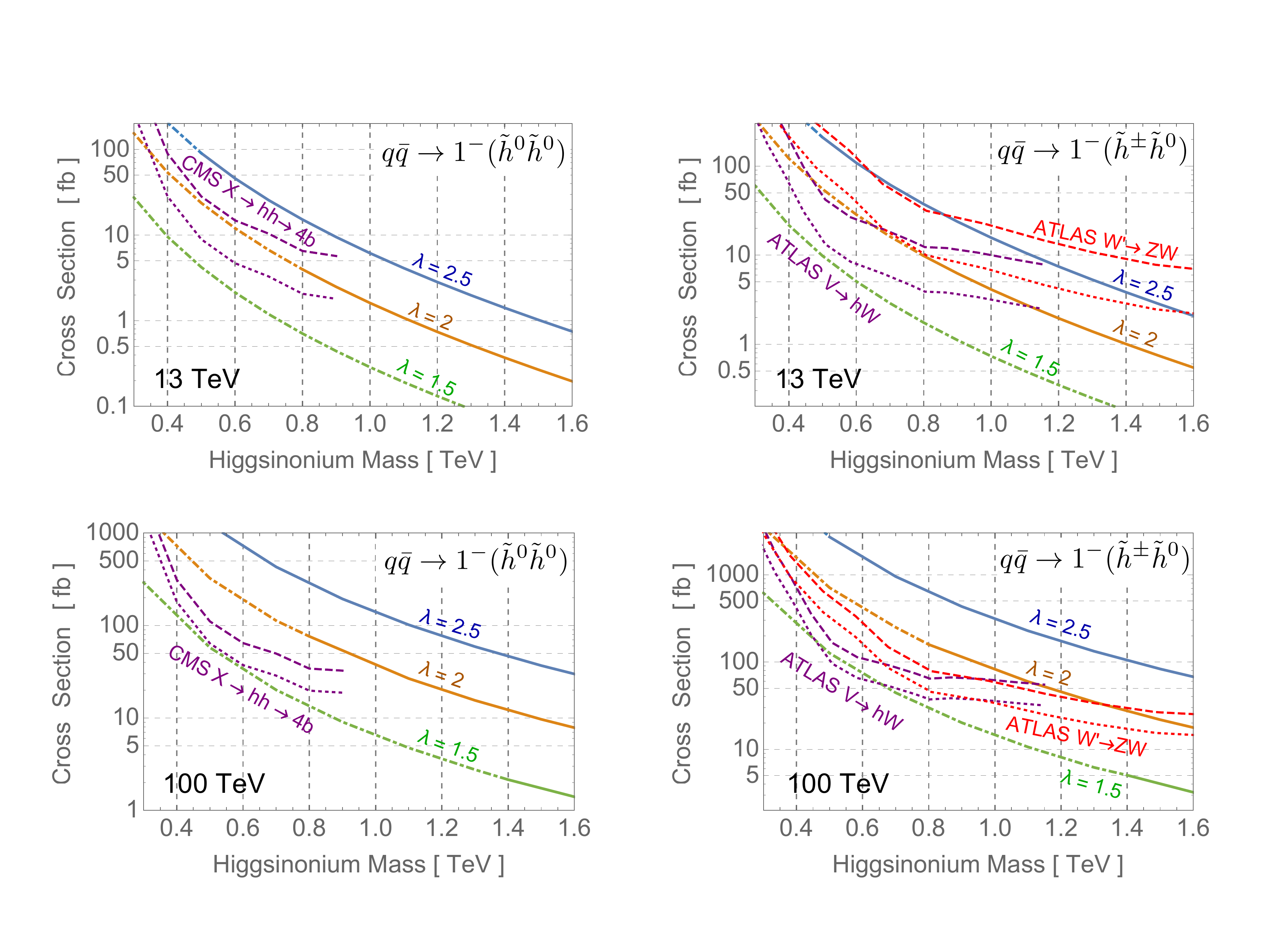}
\caption{Upper left: Higgsinonium production through $pp\to Z^*\to
(\tilde{h}^0\bar{\tilde{h}}^0)$ into $1^{-}$ bound state under
different assumptions of the $\lambda$-coupling. The solid curves
show production cross sections in which a non-vanishing mass window
Eq.~(\ref{eq:MS}) of the singlet scalar exists. The dash-dotted
extension of the curves has $m_{s}<m_h/2$, which violates the
invisible Higgs decay constraint unless a tuned coupling or an extra
singlet field is included. The purple dashed (dot-dashed) curves show
the $13$ TeV projection of the CMS resonance search $pp\to X\to
hh\to4b$ \cite{CMS:2014eda} with $300$ fb$^{-1}$ ($3$ ab$^{-1}$) of
data. The bound gives an idea of the possible future reach if the
singlet scalar has a similar mass and decay branching ratio to the
SM Higgs. Upper right: The production $q\bar{q}\to W^*\to
(\tilde{h}^{\pm}\tilde{h}^0)$ into the $1^{-}$ state. The purple
dashed (dotted) curve shows the $13$ TeV projection of the ATLAS
resonance search $pp\to V\to Wh$ \cite{Aad:2015yza} with $300$
fb$^{-1}$ ($3$ ab$^{-1}$) of data. The red dashed (dotted) curve
shows the $13$ TeV projection of the ATLAS search $pp\to W'\to WZ$
\cite{Aad:2015ufa} with $300$ fb$^{-1}$ ($3$ ab$^{-1}$) of data,
with the $Z$ branching ratio rescaled to the singlet scalar into
$b\bar{b}$. The $ZW$ bound can be used to compare with the scenario
when the singlet mass similar to $Z$, and the efficiency of the
$Z\to$jets tagging is similar for the singlet scalar decay. Lower
plots: The same plots for a $100$ TeV hadron collider. The dashed
(dotted) curves correspond to projections for $1$ ab$^{-1}$ ($3$ ab$^{-1}$) of
data.}\label{fig:product}
\end{figure*}

The production of neutral bound state goes through the SM $Z$ in Fig.~\ref{fig:diagmHiggs} left.
\begin{equation}
q\bar{q}\to Z^*\to\tilde{h}\bar{\tilde{h}}.
\end{equation}
The parton level amplitude from the vector mediation is written as
\begin{equation}
\mathcal{M}_{\tilde{h}\tilde{h}\to q\bar{q}}=\frac{g_{V\tilde{h}}}{q^2-m_Z^2}\left(\bar{v}(k')\gamma^{\mu}u(k)\right)\left(\bar{v}(p')\Gamma_{\mu}u(p)\right),
\end{equation}
where $q^2\simeq M^2=(2m_{\tilde{h}})^2$ is the bound state mass, 
and $(k',k)$ ($(p',p)$) are the Higgsino (quark) momenta. The vector coupling of Higgsino
has a coupling $g_{V\tilde{h}}$ and the latter forms a $1^-$ bound
state, while the coupling between $Z$ to quarks $\Gamma_{\mu}\equiv
g_{Vq}\,\gamma_{\mu}+g_{Aq}\,\gamma^5\gamma_{\mu}$ carries both the
vector and axial-vector components. Following the discussion in
\cite{Peskin:1995ev}, we calculate the amplitude treating quarks in
the relativistic limit.
We sum the different spin
configurations that match the total angular momentum $J=1$ and take
the average of the three $1^{-}$ polarizations for the bound state
decay rate
\begin{eqnarray}
\Gamma_{{\bf 1^{-}}\to q\bar{q}}=\frac{16\pi\,\alpha_{V\tilde{h}}(\alpha_{V\tilde{q}}+\alpha_{A\tilde{q}})|\psi(0)|^2M^2}{\left(M^2-m_Z^2\right)^2},
\end{eqnarray}
where $\lambda$ is the Yukawa coupling in Eq.~(\ref{eq:Yukawa}). Using Eq.~(\ref{eq:BScross}), we obtain the production cross section
\begin{eqnarray}\label{eq:1mmproduct}
\sigma^{NN}_{\bf{1^-}}&=&\frac{\pi^2\,\zeta(3)\,\alpha_{\lambda}^3\,\alpha_{V\tilde{h}}(\alpha_{V\tilde{q}}+\alpha_{A\tilde{q}})\,M^4}{3\,s\,(M^2-m_Z^2)^2}\,\mathcal{L}_{q\bar{q}} \nonumber \\
&=&\frac{\pi^2\,\zeta(3)\,\alpha_{\lambda}^3\,\alpha_{Z\tilde{h}}\,M^4}{3\,s\,(M^2-m_Z^2)^2}\\&\times&\left[\sum_{q}\alpha_{Zq_{V,A}}\int_{M^2/s}^1\frac{dx}{x}f_q(x)\,f_{\bar{q}}(\frac{M^2}{x\,s})+(q\leftrightarrow\bar{q})\right].\nonumber
\end{eqnarray}
The result also includes the color factor $N_c=3$. 
The production cross sections at the $13$ TeV LHC and $100$ TeV
collider are shown in Fig.~\ref{fig:product} (left). Comparing to
the Higgsino production rate at the LHC, there is $\sim 20\%$
probability that the produced Higgsinos form a $1^{-}$ bound state
when $\lambda=2$.

For the charged and neutral Higgsino production (Fig.~\ref{fig:diagmHiggs} right),
\begin{equation}
pp\to W^{\pm*}\to\tilde{h}^0\tilde{h}^{\pm},
\end{equation}
the electrically charged $1^{-}$ state is produced with a cross section
\begin{eqnarray}
\sigma^{NC}_{{\bf
1^{-}}}&=&\frac{\pi^2\,\zeta(3)\,\alpha_{\lambda}^3\,\alpha_{W}^2}{3\,s}\frac{M^4}{(M^2-m_W^2)^2}\\&\times&\left[\sum_{q}\int_{M^2/s}^1\frac{dx}{x}f_{q_u}(x)\,f_{\bar{q}_d}(\frac{M^2}{x\,s})+(q_u\leftrightarrow\bar{q}_d)\right].\nonumber
\end{eqnarray}
The result is shown in Fig.~\ref{fig:product} (right) for the $13$ TeV LHC and the $100$ TeV collider. Comparing to
the total production of the neutral and charged Higgsinos, about
$20\%$ of the events will form the $1^{-}$ bound state.

\subsection{Annihilation Decay channels}
Here we consider the possible decay channels of the bound states
\footnote{Here we assume that the mass splitting between Higgisinos
is small so that the neutralino and chargino can be treated as
stable particles before they find each other and annihilate.}. Let
us first consider the bound state formed by two neutralinos. As
discussed previously, the singlet scalar $s$ needs to be much
lighter than $m_{\tilde \chi}$. Since $\lambda$ is large, the
annihilation is dominantly to two $s$ scalars. The singlet scalar
generically mixes with the Higgs. If its mass is small, its dominant
decay channel is to $b$ quarks. If $s$ is heavier than 140 GeV,
which means the Higgsino bound states are very heavy according to
Eq.~(\ref{eq:MS}), the dominant decay is through the di-boson
channel. Similar arguments can be applied to the neutral bound state
formed by two charginos. Thus for small $m_s$, the dominant search
channel is 4 $b$-jet events with two paired resonance, and 4 $b$
jets together form a resonance of the bound state. For a heavier
$s$, the dominant channel is the 4-boson event with two paired
resonances, also with a heavier resonance from all objects.

It is important to know whether the heavy resonance is too broad to
search at the LHC. To estimate the width of the bound state, we
approximate the scattering matrix of two stationary neutralinos
annihilating to 2 scalars
\begin{eqnarray}
 \label{eq:BoundAnnM2}
|\mathcal {M}(2\tilde N \rightarrow 2s)|^2\sim \lambda^4
\end{eqnarray}
by  assuming $m_\phi$ to be much smaller than the typical energy
scale of the process, which is the mass of the bound state. Also, we
drop all angular dependence since they only contribute as $O(1)$
corrections after phase space integral. In order to convert this
2-to-2 scattetring matrix to the width of bound state, we need to
combine $|\mathcal {M}(2\tilde N \rightarrow 2s)|^2$ with the
wavefunction, which gives
\begin{eqnarray}
 \label{eq:BoundWidth}
\Gamma(B\rightarrow2s)\sim\frac{\lambda^4}{\pi}\frac{|\psi(0)|^2}{M_B^2}\sim\frac{\alpha_\lambda^5}{4}M_B
\end{eqnarray}
Now we see that the width of the bound state can be much smaller
than the mass of the bound state. For example, if the bound state is
$1$ TeV with $\lambda=2$, the width is $\simeq1$ GeV. Thus we can
safely treat the heavy resonance as a narrow width particle as long
as $\lambda$ is not too big.

To estimate how well future resonance searches can constrain the
annihilation decay, we adopt the bound from the CMS search of a
heavy resonance $X$ decaying into $hh\to 4b$'s \cite{CMS:2014eda} by
focusing on the case when the singlet scalar has a similar mass and
decay branching ratio to the SM Higgs. To compare to bounds with
future searches, we rescale the expected cross section bound from
the existing $8$ TeV search according to the parton distribution
function (PDF). We describe the details in
Appendix~\ref{app_Vrojection}. Projecting the search to the $13$ TeV
search with $300$ fb$^{-1}$ ($3$ ab$^{-1}$) of data, the upper bound
on the resonance search with $95\%$ CL is shown in the purple dashed
(dotted) curve in the upper left plot in Fig.~\ref{fig:product}.
Even with no further improvement on the search designed for the
scalars, our projection has shown the possibility of reaching the
bound state with $\lambda=2$ coupling. The bounds are only applied
to $M<900$ GeV due to the kinematic limit in the $8$ TeV search.
Higher mass bins lack statistics, and our simple $\sqrt{N}$
rescaling of the uncertainty may fail. For the future projection,
the bound for mass above $900$ GeV should be better than the lower
mass region. We leave careful study of collider constraints for
future work.

Compared to the neutralino-neutralino or chargino-chargino pair
production, the cross section of the chargino-neutralino pair
production at $13$ TeV is about 5 times larger when
$m_{\tilde{h}}>400$
 GeV. Beside the larger production cross section, the annihilation channel for the charged bound
state is also different from the neutral bound state. Due to charge
conservation, the dominant decay channel is $(W^{\pm}+s)$. Thus the
dominant signal channel is either $(W^{\pm}+2b)$ or 3-boson,
depending on $m_s$. Similar to the neutral bound state scenario,
there can be a heavy resonance. Compared to the neutral bound state
case, the width of the charged bound state is smaller because at one
of the coupling vertices $\lambda$ is replaced by the $W^{\pm}$
coupling.

Similar to the neutral bound state, we show a projection of the
ATLAS vector resonance search $V\to Wh$ \cite{Aad:2015yza} at $13$
TeV with $300$ fb$^{-1}$ ($3$ ab$^{-1}$) in the red dashed (dotted)
curve of Fig.~\ref{fig:product}(right), assuming the singlet scalar
has the same mass and decay branching ratios as the SM Higgs. To
cover the higher mass region, we also show the projected bound on
the $W'\to WZ\to\ell\nu j j$ search at ATLAS \cite{Aad:2015ufa} in
the purple dashed (dotted) curve for $13$ TeV with $300$ fb$^{-1}$
($3$ ab$^{-1}$) of data by rescaling the $Z\to$ jets branching ratio
into the singlet to $b\bar{b}$. If the tagging efficiencies between
the $s$ and $Z$ are not too different, these curves give an estimate
of the future search reach.

It is interesting to compare the reach of the Higgsinonium search to
the typical missing energy study. In \cite{Low:2014cba}, the authors
estimate the future bound on the Higgsino mass from the
monojet$+$MET search, assuming the Higgsino to be the lightest SUSY
particle. The $95\%$ CL constraint from the $14$ TeV study with $3$
ab$^{-1}$ of data is $m_{\tilde h}\gsim 200$ GeV, and from the $100$
TeV collider is $m_{\tilde h}\gsim 700$ GeV. When $\lambda\gsim 2$,
our bounds in Fig.~\ref{fig:product} can be better than these
results.

\section{The SIDM bound state}\label{sec:sidm}
The bound state annihilation decay may also exist in DM models with
a strong self-interaction. An important motivation for the SIDM
model is the possibility of solving anomalies in small scale
structures, including the too big to fail problem and the
disagreement of the core/cusp halo structure obtained between
observation and N-body simulations \cite{ simulation1, simulation2,
simulation3, Tulin:2013teo}. When considering the structure of dwarf
DM halos, the self-interaction with $\sigma_T/m_{\chi}\sim 0.5-50$
cm$^2/g$ on dwarf scales can produce smooth core density in dwarf
galaxies in accordance with observations \cite{coreformat,
sidmsimulat}.

In this work, we consider light mediator models that can generate a
cross section favored by DM profile measurements in dwarf galaxies.
We study the annihilation decay of the SIDM particles at colliders.
We assume DM, $\chi$, is fermionic. Its self-interaction is induced
by a light scalar mediator $\phi$ through the Yukawa coupling
$\lambda_{\chi}\bar{\chi}\phi\chi+$ h.c., or a vector mediator $A'$
through a gauge coupling $-i\lambda_{\chi}\bar{\chi}\slashed
A'\chi$. The self-scattering $\chi\bar{\chi}\to\chi\bar{\chi}$ has a
t-channel enhancement in the non-relativistic limit when mediator mass is lower than the momentum transfer.

In order to form DM bound states at colliders, we need the mediator
wavelength to be longer than the Bohr radius,
$\alpha_{\chi}m_{\chi}/m_{med}\gsim 2$
($\alpha_{\chi}\equiv\lambda_{\chi}^2/4\pi$). Born approximation
does not apply to DM self scattering any more when
$\alpha_{\chi}m_{\chi}/m_{med}\gsim 1$. Thus we have to take into
consideration non-perturbative effects.

Further, as we will discuss in a later section, if $m_{med}\ll 10$
MeV, the decay length of the mediators from the DM annihilation is
generically too long for collider searches. We thus limit the
discussion to mediator masses $m_{med}\gsim 10$ MeV. DM particles
with mass around $\mathcal{O}(1-100)$ GeV is our focus from a
collider point of view, thus we have $m_{\chi}v/m_{med}\lsim 1$,
where $v\sim 10^{-4}$ is the viral velocity of dwarf galaxies. Under
this choice of parameters, quasi-bound states of DM can form, in
which case the quantum mechanical resonances and anti-resonances
emerge for the SIDM interaction. The analytical approximation
obtained in this regime is written as (for the full expression,
please see \cite{Tulin:2013teo})
\begin{equation}\label{eq:sidmscatt}
\sigma_T=\frac{16\pi}{m_{\chi}^2v^2}\sin^2\delta_0.
\end{equation}
For an attractive force, the resonance effect makes
$\sin\delta_0\to1$ when $\alpha_{\chi}m_{\chi}/1.6 m_{med}=n^2$,
$n=1,2,3,...$ in the small velocity limit. On the other hand,
anti-resonance, with vanishing $s$-wave cross section, happens when
$\sin\delta_0\to0$, i.e. $n\approx1.69,\,2.75,\,3.78,...$. If the
force is repulsive, which happens in an asymmetric DM model with a
dark photon mediator, there is no resonance or anti-resonance
effect, and the cross section is calculated by the full expression
in Eq.~(\ref{eq:sidmscatt}). For the DM mass and couplings we study
below, the self-interacting cross section satisfies the Bullet
Cluster and cluster shape constraints with the typical velocity
$v\sim 10^{-2}$ and $\sigma_T/m_{\chi}\lsim 1$ cm$^2/g$
\cite{Randall:2007ph,sidmsimulat}.

For a given value of $(m_{\chi},\alpha_{\chi})$, we can obtain the
mediator mass $m_{med}$ that gives the right scattering cross
section ($0.5-50$ cm$^2/$g) by solving Eq.~(\ref{eq:sidmscatt}). The
solution is not unique due to the finite range of the scattering
cross section and the resonance/anti-resonance behavior in the
attractive case. When the $m_{\chi}$ is heavy (light), the solution
of the mediator mass in the attractive case is closer to the
(anti-)resonance region. We make sure the mediator mass as the
solution also satisfies $m_{med}<\alpha_{\chi}m_{\chi}/2$ required
by the bound state production. In the following study, we make sure
each choice of $(m_{\chi},\alpha_{\chi})$ has a corresponding
$m_{med}$ satisfying the above constraints. However, since the bound
state production is insensitive to the mediator mass, we keep the
value of $m_{med}$ implicit when showing the results.

In a thermal relic DM scenario, the size of $\alpha_{\chi}$ has an
upper bound from the DM density, which limits the collider
production of the DM bound states. However, as this bound can be
avoided in various scenarios, such as the context of asymmetric DM
or non-thermal production, we will not take it into account for the
study.
\subsection{Darkonium Production}
In order to study the bound state production rate, we parameterize
the DM-SM interaction by effective operators
\begin{equation}
0^{-}:\,\, \frac{m_q(\bar{q}\gamma^5q)(\bar{\chi}\gamma^5\chi)}{v_h\,M_*^2},\qquad 1^{-}:\,\,\frac{(\bar{q}\gamma^{\mu}\gamma^5q)(\bar{\chi}\gamma_{\mu}\chi)}{M_*^2}.
\end{equation}
Here $v_h=174$ GeV, and the quark mass in the pseudo scalar coupling can come from a
straightforward UV completion in which the chiral symmetry breaking
induces a Yukawa coupling insertion. The $\gamma^5$ in the vector
mediation causes velocity suppression in DM direct detection
experiments. Also, the scattering with DM and nucleus is
spin-dependent. Thus this operator is less constrained. In contrast
to the missing energy search at high energy colliders, the use of
effective operators in bound state production is well justified. The
center of mass energy is fixed to be around $2m_{\chi}$. This is
much lower than $4\pi M_*$, which can be probed in collider
searches.

\begin{figure}
\center
\includegraphics[width=8cm]{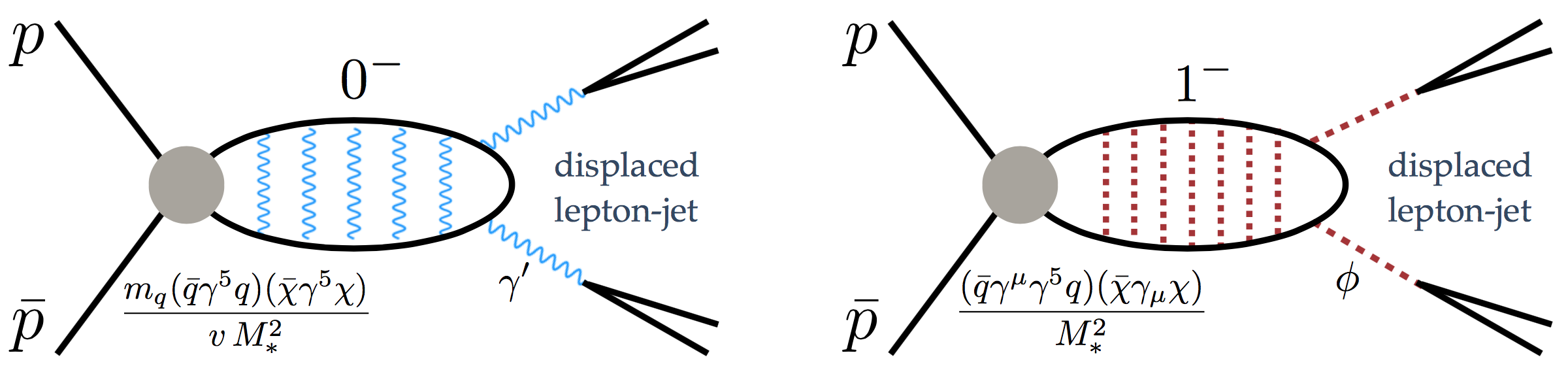}
\caption{Darkonium production and decay at the LHC.}\label{fig:diagmSIDM}
\end{figure}
\begin{figure*}
\begin{center}
\includegraphics[width=8.2cm]{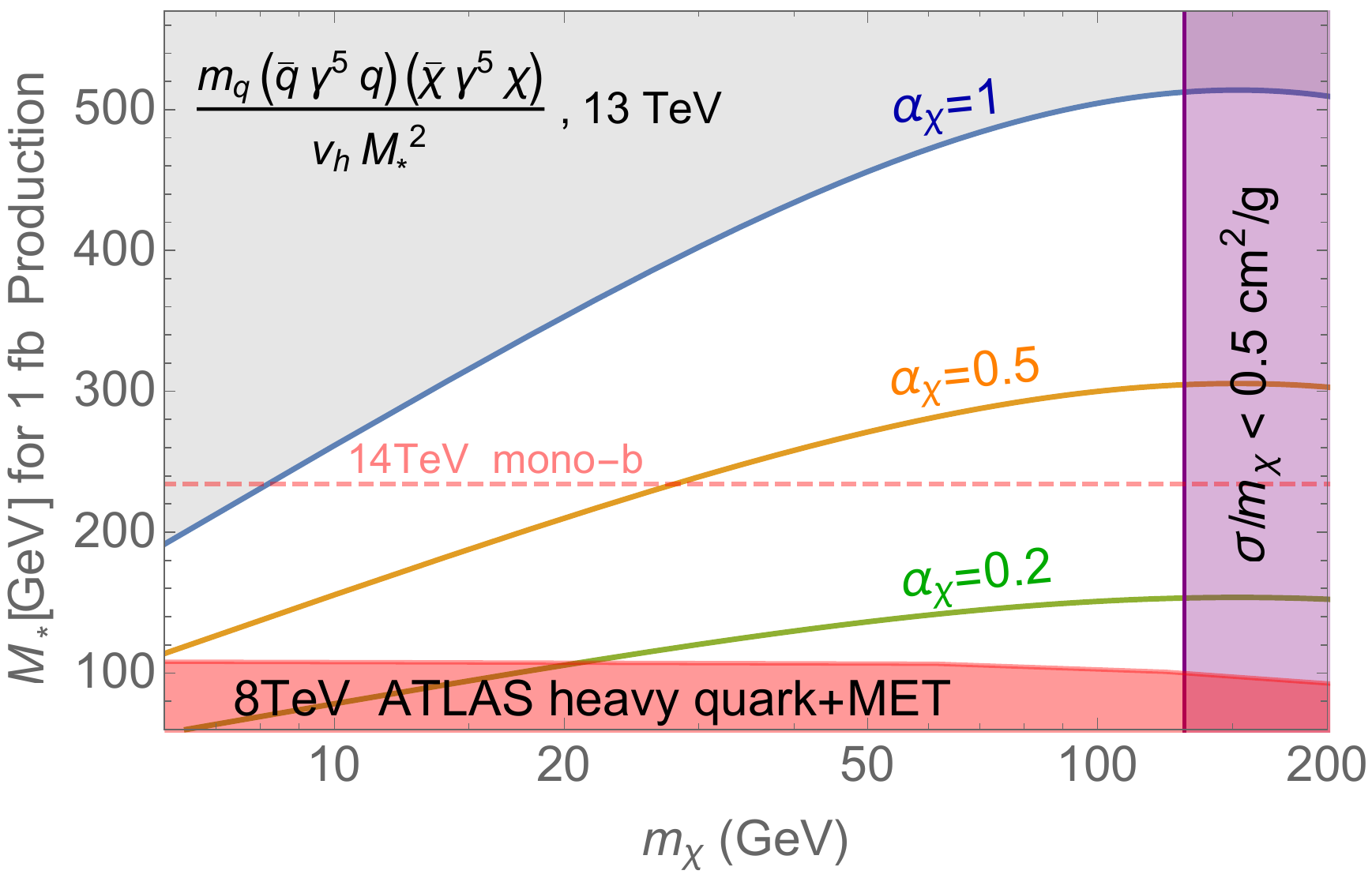}\qquad\includegraphics[width=7.9cm]{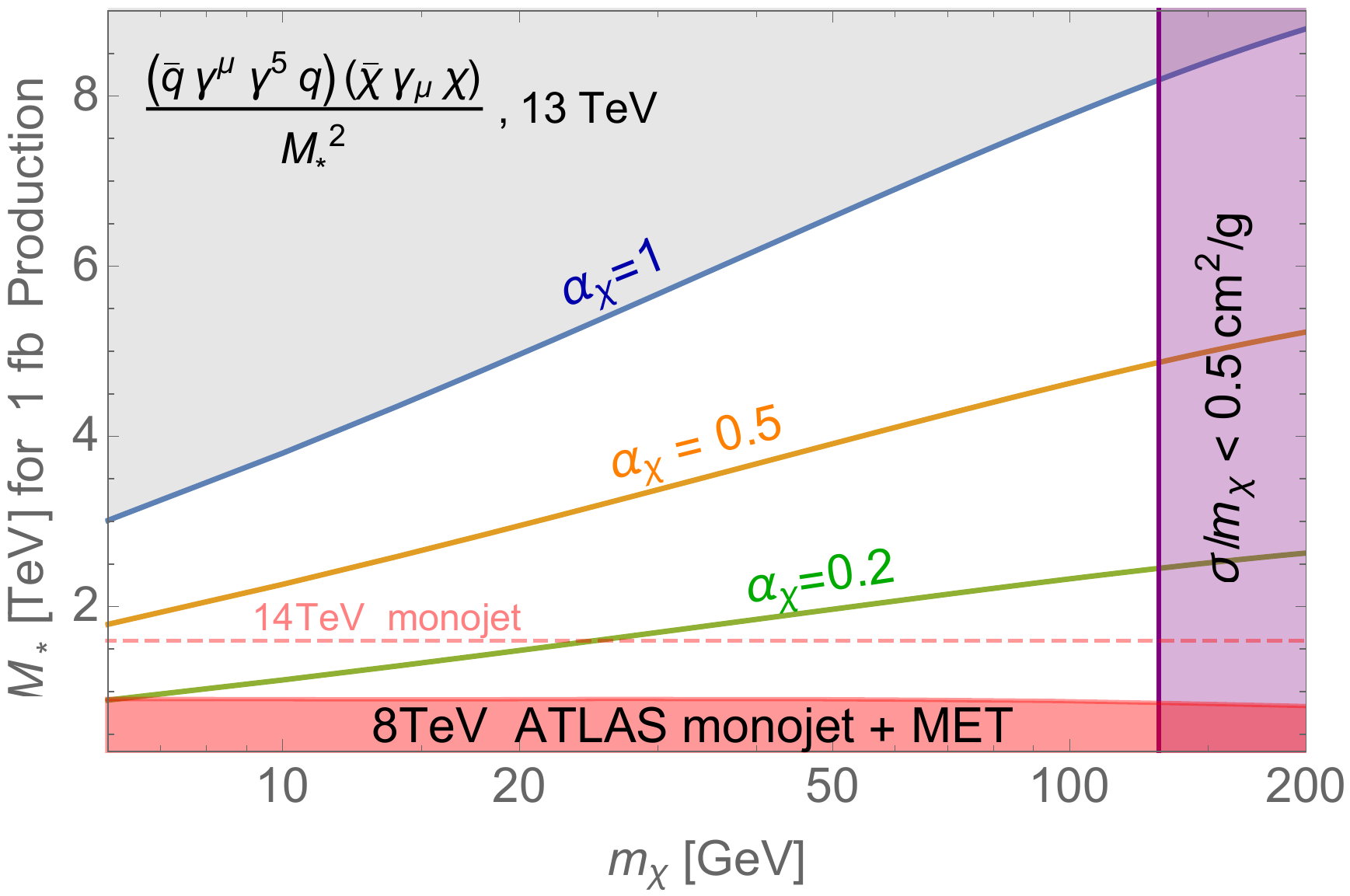}
\caption{Left: The mediation scale $M_*$ for having a $1$ fb $0^-$
bound state production through the pseudo-scalar mediation at $13$
TeV LHC. The red-shaded region shows the $95\%$ CL exclusion bound
from the $8$ TeV ATLAS heavy quark$+$MET search \cite{Aad:2014vea}.
The dashed curve shows the $95\%$ CL bound from the mono-$b$
search estimated in \cite{Artoni:2013zba}, assuming $300$ fb$^{-1}$
of data at $14$ TeV LHC. The purple-shaded region has the DM
scattering $\sigma/m_{\chi}<0.5$cm$^2/$g at dwarf galaxies (assume
$v=10^{-4}$), assuming $m_{med}$ matches the size of $(m_{\chi},\alpha_{\chi})$ that gives $\sin\delta_0=1$ in Eq.~(\ref{eq:sidmscatt}). The
gray-shaded region corresponds to $\alpha_{\chi}>1$, for which the
non-relativistic calculation of the bound state production fails.
Right: The $1$ fb region of the $1^{-}$ state through the vector
coupling. The red-shaded region corresponds to the $95\%$ CL
exclusion bound from the $8$ TeV ATLAS monojet search
\cite{Aad:2015zva}. The dashed curve shows the $95\%$ CL
bound from the mono-jet search estimated in \cite{Zhou:2013raa},
assuming $300$ fb$^{-1}$ of data at $14$ TeV LHC.} \label{fig:SIDM2}
\end{center}
\end{figure*}

For the $0^{-}$ state from the quark production, the decay rate
from the bound state into quarks is written as
\begin{equation}\label{eq:0mpdecay}
\Gamma_{0^{-}\to q\bar{q}}=\frac{3}{\pi M^2}\left(\frac{M}{M_*}\right)^4\left(\frac{m_q}{v_h}\right)^2|\psi(0)|^2.
\end{equation}
where $M$ is the mass of the bound state. The result includes a
color factor and a summation of the correct spin configurations that
match the $J=0$ state. Using Eq.~(\ref{eq:BScross}), the production
cross section is written as
\begin{eqnarray}
\sigma_{
0^{-}}&=&\frac{\zeta(3)\,\alpha_{\chi}^3}{48\,s}\left(\frac{M}{M_*}\right)^4\\&\times&\left[\sum_{q}\left(\frac{m_{q}}{v_h}\right)^2\int_{M^2/s}^1\frac{dx}{x}f_q(x)\,f_{\bar{q}}(\frac{M^2}{x\,s})+(q\leftrightarrow\bar{q})\right].\nonumber
\end{eqnarray}
In Fig.~\ref{fig:SIDM2} we show the region of mediation scale $M_*$
that gives at least 1 fb bound state production rate with different
choices of the SIDM coupling. The smaller $M_*$ region is excluded
by the ATLAS heavy quark search at the LHC Run 1. For this operator,
the $b$-quark dominates the production. When the center of mass
energy is larger for heavier DM production, $b$-quark PDF decreases
faster compared to that of light quarks. This makes the bound on
$0^{-}$ weaken faster than the production channels, which are
dominantly through light quarks, e.g. $1^{-}$ as discussed later. We
require $\alpha_\chi<1$ for perturbation calculation, which implies
that the parton in the bound state is non-relativistic.

Similar to the Higgsino case, the bound state production of $1^{-}$
through the axial-vector mediated process can be obtained by
Eq.~(\ref{eq:1mmproduct}) with $g_V=1$ and $M_V=M_*\gg M$
\begin{eqnarray}
\sigma_{1^{-}}&=&\frac{\zeta(3)\alpha_{\chi}^3}{48\,s}\left(\frac{M}{M_*}\right)^4\\ &\times& \left[\sum_{q}\int_{M^2/s}^1\frac{dx}{x}f_{q}(x)\,f_{\bar{q}}(\frac{M^2}{x\,s})+(q\leftrightarrow\bar{q})\right].\nonumber
\end{eqnarray}
In the right plot of Fig.~\ref{fig:SIDM2}, we show the region of
mediation scale $M_*$ which gives at least 1 fb bound state
production cross section with various choices of the SIDM coupling.

\subsection{Annihilation Decay Channels}
With large self couplings, the SIDM bound states prefer to decay
into light mediators rather than the SM quarks. Instead of surveying a comprehensive list of possible mediators, we focus on a few illustrative examples. For vector bound
state $1^{-}$,  we consider the case where the mediator is a scalar.
The decay of the bound state into two light scalars can be
characterized by a derivative coupling
$V_{1^-}^{\mu}\phi^*i\overset{\leftrightarrow}{\partial_{\mu}}\phi$
\footnote{The derivatives in the operator do not cause any
suppression of decay width, since the typical momentum of decay
products is comparable to DM mass. One can also consider the
scenario where the mediator is a vector boson, such as light dark
photons. However, the decay in that case suffers additional
suppressions from the dark photon mass due to Landau-Yang theorem.
Since the collider study is similar to the scalar channel, we do not
consider it in this work.}. This decay rate is of order
$\sim\alpha_{\chi}^5M/8\pi$, which gives a prompt $1^{-}\to
\phi\phi^*$ decay.

Generically, the force mediators are not stabilized by a symmetry
(for exceptions, see \cite{Curtin:2013qsa,Curtin:2014afa}) and can
decay into SM particles. For the scalar mediator $\phi$, we 
parameterize the decay using the effective coupling
$\hat{y}_{ij}\phi
\bar{L}_iHE_j/\Lambda\supset\epsilon_{\phi,ij}\phi\bar{\ell}_i\ell_j$,
where $\hat{y}_{ij}$ is aligned to the SM Yukawa coupling. The
scalar can also couple to SM through the Higgs mixing, but the
coupling is generically more suppressed due to constraints from the
Higgs coupling measurement \cite{ATLAS-CONF-2015-044}, as well as the small
Yukawa coupling to the light SM fermions. If the effective coupling is generated by an EW scale mediation, and $\hat{y}_{ij}$ is indeed the SM Yukawa coupling, $\epsilon_{\phi,ee}$ can
be as small as $O(10^{-6})$. Fig.~\ref{fig:diagmSIDM} (left) shows the production
and decay of the bound state.

On the other hand, we assume the $0^{-}$ bound state decays into
dark photons. The decay of $0^{-}$ can be described by a
pseudo-scalar coupling
$\frac{i}{\Lambda'}a_{0^-}F'_{\mu\nu}\tilde{F}'^{\mu\nu}$. In the
microscopic picture, the annihilation
$\chi\bar{\chi}\to\gamma'\gamma'$ that generates the
$F'_{\mu\nu}\tilde{F}'^{\mu\nu}$ interaction needs to break the
parity. This requires the dark photon to couple chirally to the DM
fermions. Further, dark photons decay to SM through kinetic mixing
$\epsilon_{\gamma'} F_{\mu\nu}F'^{\mu\nu}$ to the normal photon.
Currently, the bound from the various dark photon searches requires
$\epsilon_{\gamma'}\lsim 10^{-3}$ for $m_{\gamma'}>10$ MeV
\cite{Curtin:2014cca}. We will study the $\gamma'$ decay with a
mixing satisfying the bound. Fig.~\ref{fig:diagmSIDM} (right) shows
the production and decay of the bound state.

We focus on the mediator decay into $e^+e^-$. As heavier mediators
open up other decay channels such as muon and pion, we leave a more
complete analysis for future work. The two types of mediator decay
have lengths
\begin{eqnarray}\label{eq:phiee}
c\tau_{\phi\to e^+e^-}&\simeq&\gamma_{\phi}\left(\frac{\epsilon_{\phi}^2\,m_{\phi}}{8\pi}\right)^{-1}\nonumber\\&\simeq& 5\,\text{cm}\times\gamma_{\phi}\left(\frac{100\,\text{MeV}}{m_{\phi}}\right)\left(\frac{10^{-6}}{\epsilon_{\phi}}\right)^2,
\\c\tau_{\gamma'\to e^+e^-}&\simeq&\gamma_{\gamma'}\left(\frac{e^2\epsilon_{\gamma'}^2\,m_{\gamma'}}{12\pi}\right)^{-1}\nonumber\\&\simeq& 0.08\,\text{mm}\times\gamma_{\gamma'}\left(\frac{100\,\text{MeV}}{m_{\gamma'}}\right)\left(\frac{10^{-4}}{\epsilon_{\gamma'}}\right)^2.
\end{eqnarray}
The boost factor $\gamma_{\phi,\gamma'}\simeq
m_{\chi}/m_{\phi,\gamma'}$ can be larger than $10^2$, which provides
a good chance to observe displaced decays. However, the large boost
also corresponds to a small opening angle between $e^+e^-$, which
makes searches relying on reconstructing the displaced vertices (DV)
difficult. Even if the magnetic field can eventually open up the
$e^+e^-$ angle, multi-scatterings of the $e^+e^-$ inside the tracker
and ECAL can still limit the precision. Further study including the
detector performance is then necessary for the standard DV search.

One displaced search that cares less about opening angle is the
displaced lepton jet (DLJ) search. When the decay length is long
enough, a sizable fraction of the decays can happen in the HCAL,
making a ``jet" that does not show up in the ECAL and the tracker.

To estimate the bounds on the DM bound state production, we adopt
the cuts used in the $8$ TeV ATLAS search for the displaced LJ
\cite{Aad:2014yea} by requiring two $e^+e^-$ jets produced in the
HCAL with $2.0<r_{\text{DLJ}}<3.6$ m, $\Delta R(e^+,e^-)\leq 0.5$,
$p_T(\text{LJ})>30$ GeV, and $|\eta(\ell)|<2.5$. We carry out our
simulation at the $13$ TeV LHC with $300$ fb$^{-1}$ of data. Beside
the cuts, a $25\%$ reconstruction efficiency for identifying both
lepton-jets is multiplied to get the number of signal events. This
number is close to the efficiency obtained in the ATLAS study. In
contrast to the dark photon search in \cite{Aad:2014yea}, we can
further require the total invariant mass of the DLJ's to be near the
DM bound state mass, which will further reduce the SM background
dominated by the multi-jets and cosmic ray events.

\begin{figure*}
\begin{center}
\includegraphics[width=8.2cm]{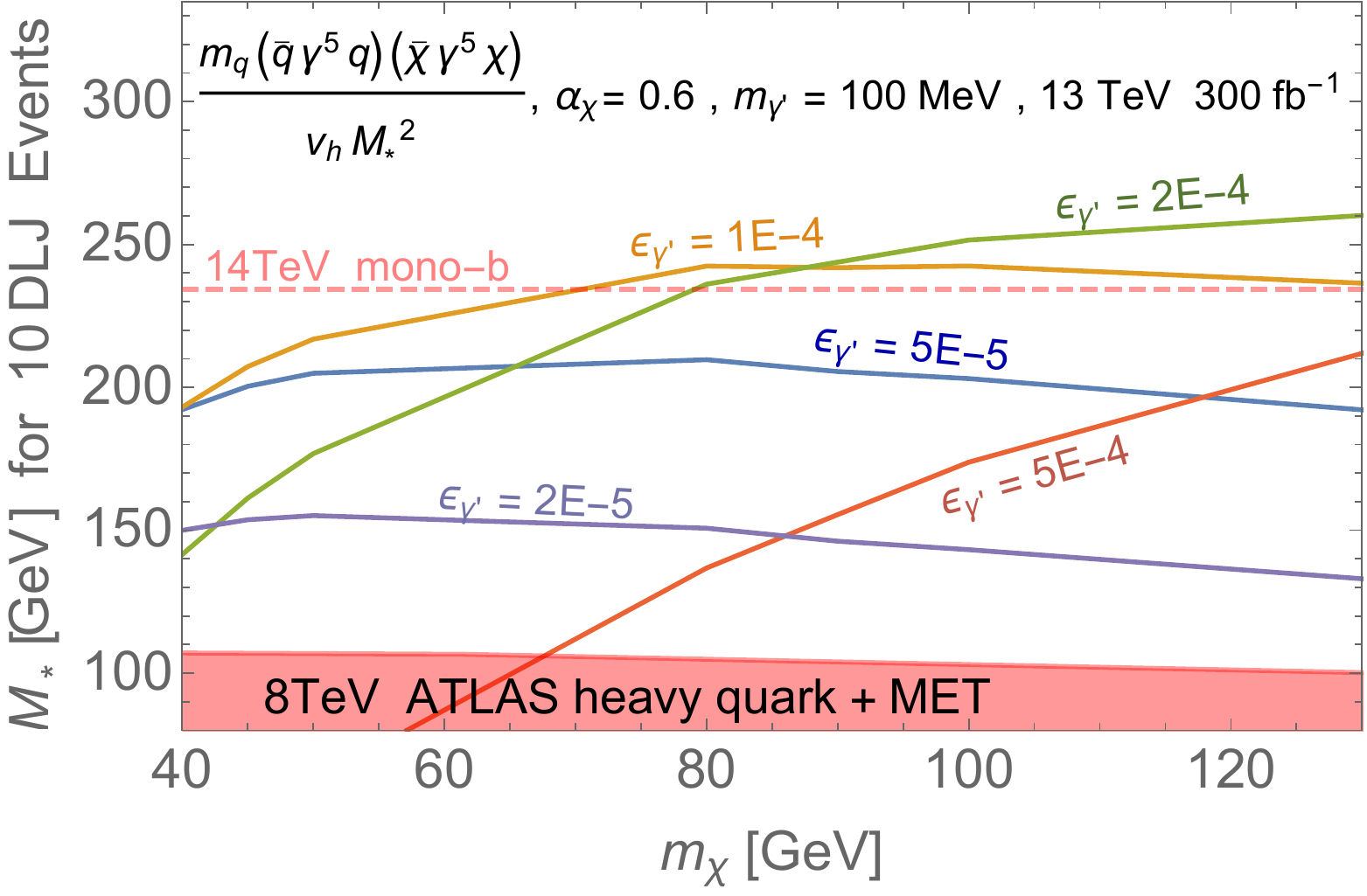}\qquad\qquad\includegraphics[width=7.85cm]{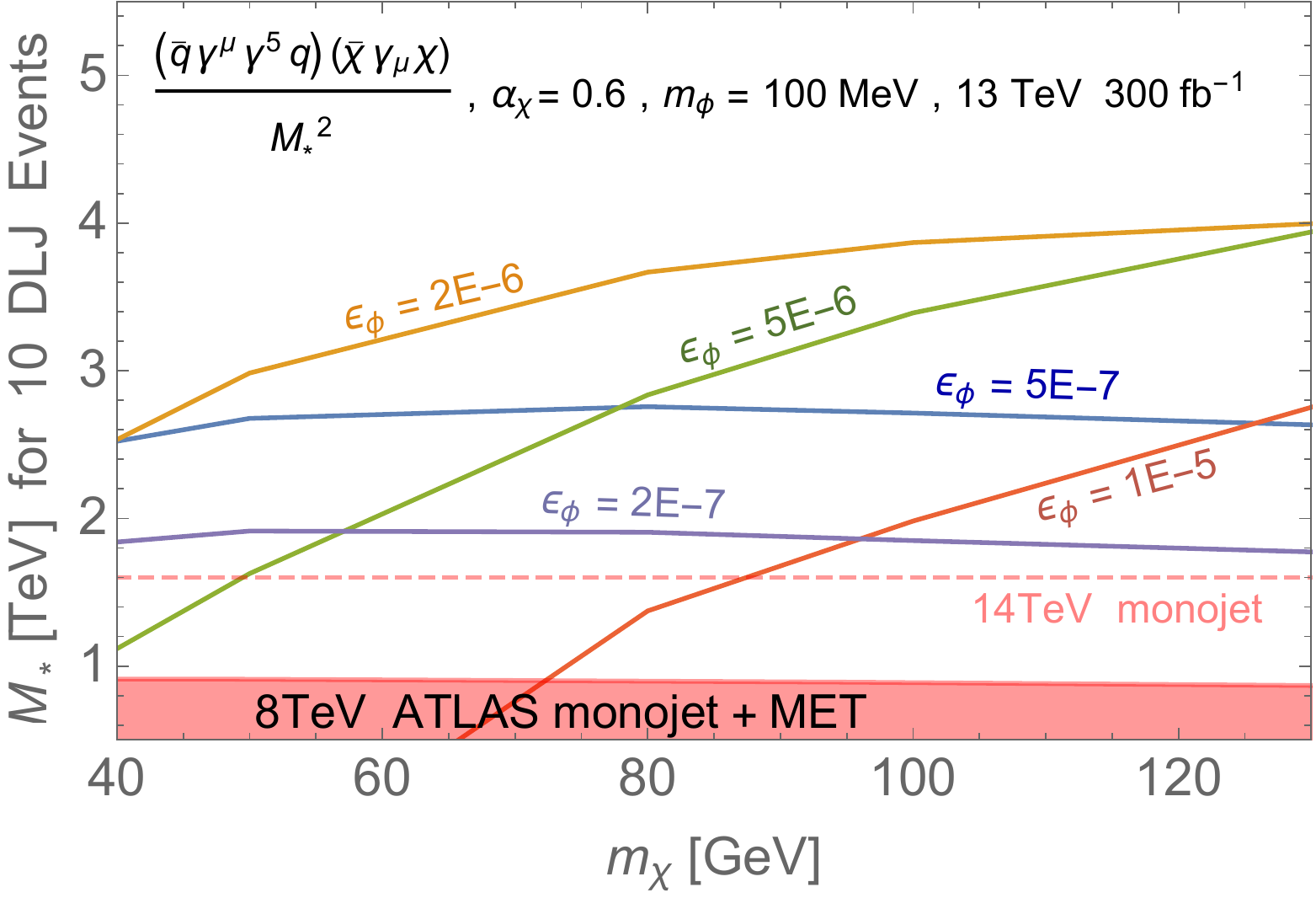}
\caption{Left: The mediation scale $M_*$ of the pseudo-scalar
operator for the $0^{-}$ bound state production. The $0^{-}$ decays
into two dark photons, which have displaced decays into $e^+e^-$ and
form displaced lepton jet (DLJ) signals. The light mediator mass is
fixed to be $m_{med}=100$ MeV, and different curves correspond to
different sizes of the mixing
$\epsilon_{\gamma'}F_{\mu\nu}F'^{\mu\nu}$. We require $10$ events
containing two identified DLJ's at the HCAL from the dark photon
decay, assuming $300$ fb$^{-1}$ of data at $13$ TeV LHC. See details
in the text for the assumed cuts and reconstruction efficiency.
Right: The mediation scale $M_*$ of the axial-vector operator for
the $1^{-}$ bound state production. The $1^{-}$ decays into two
scalar mediators, which have displaced decays into $e^+e^-$ and give
the DLJ signals. Different curves correspond to different sizes of
the coupling $\epsilon_{\phi}\phi\bar{e}e$. The current and
projected bounds from the missing energy searches are described in
Fig.~\ref{fig:SIDM2}. Two remarks: First, compared to the bound
state production rate in Fig.~\ref{fig:SIDM2}, the signal efficiency
of our DLJ study is of order $0.1-1\%$. A better searching strategy,
such as including signals in the tracker and $\mu$-chamber, may
improve the result. Moreover, the $M_*$ bound from the mono-$b$
search is much lower than the typical center of mass energy at $14$
TeV LHC. Therefore, simplified  models with light mediators will give more accurate descriptions of collider signal, and the result will depend on the assumption of mediator
coupling. Here we include results from effective operators to show
an estimated mediation scale that the missing energy searches can
reach.}\label{fig:jetDV}
\end{center}
\end{figure*}

In Fig.~\ref{fig:jetDV}, we show the estimated lower bounds on the
mediation scales in the effective operators of $0^{-}$ and $1^{-}$
production, assuming the exclusion bound requires less than $10$
observed events with $m_{\phi,\gamma'}=100$ MeV. The bound is
obtained by a parton level study using MadGraph~5
\cite{Alwall:2011uj} and model generator Feynrules~2.3
\cite{Alloul:2013bka}. In order to capture the correct energy and
angular distribution of the light mediators, we use the effective
coupling $a F'_{\mu\nu}\tilde{F}'^{\mu\nu}/\Lambda'$ to describe the
$0^{-}$ decay into dark photons, and
$\lambda_{\chi}V_{\mu}\phi^*\,\partial^{\mu}\phi$ for the $1^{-}$
decay into scalars. The decay probability within a given decay
length is calculated using the events passing the energy and angular
cuts. As one can see, the displaced LJ search can explore a wide
range of currently unconstrained mediation scales. When the mixing
angle is small (the purple, blue curves), the mediator tends to
decay outside the collider. Bounds on a lighter DM is then stronger
because it gives a smaller boost to the mediators. On the other
hand, with a larger mixing (the green, red curves), the meditator
tends to decay before it reaches ECAL. Thus a heavier DM is more
constrained.

\section{Summary}\label{sec:summary}
In this paper, we study the collider physics of bound state
production of strongly coupled particle pairs. Such scenarios are
well motivated by the large Yukawa coupling mediated by gauge
singlet or Higgs boson in $\lambda$-SUSY, as well as the
self-interacting dark matter scenario.

In $\lambda$-SUSY, the large value of $\lambda$ introduces a sizable
Yukawa coupling. The pair produced neutralino/chargino at the LHC
can form a bound state by exchanging singlet/Higgs. The annihilation
decay final states of neutralino pair or chargino pair are
dominantly the scalar mediator. If the scalar mediator is lighter
than 140 GeV, the dominant search channel is 4-b jets. These four b
jets pair up to form two lighter peaks, and all four b jets form a
heavier resonance. Otherwise, the final state has 4 $W$ bosons if 
both mediator and Higgsino are heavy. On the other hand, the bound state
formed by chargino and neutralino dominantly decay to one W boson
and one scalar mediator. When the mediator is light, one can look
for $2b+1 l+$MET or boosted hadronic W and 2 b jets. This particular
channel has not been studied in detail. If the mediator is heavy,
the final state has 3 $W$ bosons. In order to estimate the reach, we
rescale the result of heavy vector boson search at 8 TeV with
respect to parton luminosity. Note that this rescaling procedure is
only valid when statistical error dominates and the number of events
is large in the relevant bins. Thus this estimate is only valid when
the bound state mass is below TeV. If the bound state is very heavy,
the number of events in the high energy bins is quite small. A naive
rescaling tends to give a too conservative result.  A more careful
collider physics search is necessary in order to get a more precise
estimation. With the conservative estimation we carry out, we find
the 13 TeV LHC running can already probe interesting regions of
$\lambda$-SUSY, $\lambda\simeq 2$. The reach can be further improved
at a 100 TeV machine.

In SIDM, the DM self-interaction may be explained by a light force
mediator strongly coupled to DM particles. Bound state of DM can be
formed at the collider by exchanging the light force mediator. The
annihilation decay of this DM bound state will dominantly produce
two highly boosted mediators. The decay of the mediator can be
either prompt or displaced, depending on how it mixes with the SM
sector and how strong the mixing is. If the decay is prompt, one
needs to look for the signature of double bump that further reconstruct a
heavier resonance. If the decay is displaced, the signature is more
spectacular and the SM background is very small. Compared to various
mono-X+MET searches, forming bound states provides much easier
access to new physics and allows mass measurements of particles in the dark sector. We have shown that the sensitivity from the
LHC can go well beyond the reach of mono-jet searches. Further, it
is more proper to use the language of effective operators to
describe DM production in the bound state scenario than the mono-jet
scenario. The typical energy of the bound state production process
is fixed to be the mass of the bound state, which is usually below
the mediation scale that one constrains. The validity of using an
effective operator description reduces subtleties on the conclusion
one can draw from mono-X searches.

\acknowledgements We are grateful to Guido Ciapetti and Hai-Bo Yu for useful discussions. We also thank Bibhushan
Shakya for reading the draft and give important comments. Y.T. is supported by
the National Science Foundation Grant No. PHY-1315155, and by the Maryland Center
for Fundamental Physics. L.W. is supported in part by the Kavli Institute for Cosmological Physics at the University of Chicago through grant NSF PHY-1125897 and an endowment from the Kavli Foundation and its founder Fred Kavli. YZ is supported by
DE- SC0007859.
\appendix
\section{Projection of future searches}\label{app_Vrojection}
Here we discuss the projection of the $8$ TeV vector resonance
search to the $13$ TeV LHC, following the concepts described in
\cite{Chang:2014ida}. The significance of a resonance search is
dominated by the number of signals versus background near the
resonance peak. Since the number of events in a hadron machine is
dominated by the PDF, we can project the future bound by rescaling
the PDF at a higher energy. Taking the current cross section
constraints of the new signal as a function of the invariant mass,
$\sigma^8_S(M)$, we obtain the bound at $13$ TeV by solving
\begin{eqnarray}
\frac{\sigma^{8}_S(M)\,L_{8}}{\sqrt{(\sigma^{8}_B(M)\,L_{8}\,\epsilon_8)^2+\sigma^{8}_B(M)\,L_{8}}}\nonumber\\=\frac{\sigma^{13}_S(M)\,L_{13}}{\sqrt{(\sigma^{13}_B(M)\,L_{13}\,\epsilon_{13})^2+\sigma^{13}_B(M)\,L_{13}}}.
\end{eqnarray}
Here $L_{8,13}$ are the integrated luminosity at the $8$ and $13$
TeV searches. $\sigma^{8,13}_B(M)$ is the SM background near the resonance
peak, which we assume to be rescaled by the PDF with
\begin{eqnarray}
\sigma^8_B(M)/\sigma^{13}_B(M)\sim\mathcal{L}_{ij}(M,s_{8})/\mathcal{L}_{ij}(M,s_{13}),\nonumber\\\mathcal{L}_{ij}=\frac{M^2}{s}\int^1_{M^2/s}\frac{dx}{x}f_i(x)f_{j}(\frac{M^2}{xs}).
\end{eqnarray}
Assuming the percentage systematic uncertainty $\epsilon_{13}$ at
$13$ TeV will be scaled down by the increasing statistics of the
relevant control sample events,
$\epsilon_{13}\approx\epsilon_{8}\sqrt{\sigma^{8}_B(M)\,L_{8}/\sigma^{13}_B(M)\,L_{13}}$,
the projected bound is written into
\begin{equation}
\sigma^{13}_S(M)\approx\sqrt{\frac{L_8}{L_{13}}}\,\sqrt{\frac{\sigma^{13}_B(M)}{\sigma^8_B(M)}}\,\sigma^8_S(M).
\end{equation}
In more detail:
\begin{itemize}
\item CMS $X\to hh\to 4b$ \cite{CMS:2014eda}: the $8$ TeV search uses $17.9$ fb$^{-1}$ of data, with the dominant background given by the multi-jet events. We use the gluon PDF to rescale the background cross section.
\item ATLAS $V\to Wh\to$ $\ell\nu b\bar{b}$ \cite{Aad:2015yza}:  the $8$ TeV search uses $20.3$ fb$^{-1}$ of data, with the dominant background given by $t\bar{t}$. We use the gluon PDF to rescale the background cross section.
\item ATLAS $W'\to WZ\to\ell \nu jj$ \cite{Aad:2015ufa}: the $8$ TeV search uses $20.3$ fb$^{-1}$ of data, with the dominant background given by the $W+$jets events. We use the quark PDF to rescale the background cross section. We also multiply the branching ratio of $Z\to$jets into the result and reproduce the actual cross section being constrained in the search.
\end{itemize}
\bibliography{./DMreference}

\begin{thebibliography}{68}%
\makeatletter
\providecommand \@ifxundefined [1]{%
 \@ifx{#1\undefined}
}%
\providecommand \@ifnum [1]{%
 \ifnum #1\expandafter \@firstoftwo
 \else \expandafter \@secondoftwo
 \fi
}%
\providecommand \@ifx [1]{%
 \ifx #1\expandafter \@firstoftwo
 \else \expandafter \@secondoftwo
 \fi
}%
\providecommand \natexlab [1]{#1}%
\providecommand \enquote  [1]{``#1''}%
\providecommand \bibnamefont  [1]{#1}%
\providecommand \bibfnamefont [1]{#1}%
\providecommand \citenamefont [1]{#1}%
\providecommand \href@noop [0]{\@secondoftwo}%
\providecommand \href [0]{\begingroup \@sanitize@url \@href}%
\providecommand \@href[1]{\@@startlink{#1}\@@href}%
\providecommand \@@href[1]{\endgroup#1\@@endlink}%
\providecommand \@sanitize@url [0]{\catcode `\\12\catcode `\$12\catcode
  `\&12\catcode `\#12\catcode `\^12\catcode `\_12\catcode `\%12\relax}%
\providecommand \@@startlink[1]{}%
\providecommand \@@endlink[0]{}%
\providecommand \url  [0]{\begingroup\@sanitize@url \@url }%
\providecommand \@url [1]{\endgroup\@href {#1}{\urlprefix }}%
\providecommand \urlprefix  [0]{URL }%
\providecommand \Eprint [0]{\href }%
\providecommand \doibase [0]{http://dx.doi.org/}%
\providecommand \selectlanguage [0]{\@gobble}%
\providecommand \bibinfo  [0]{\@secondoftwo}%
\providecommand \bibfield  [0]{\@secondoftwo}%
\providecommand \translation [1]{[#1]}%
\providecommand \BibitemOpen [0]{}%
\providecommand \bibitemStop [0]{}%
\providecommand \bibitemNoStop [0]{.\EOS\space}%
\providecommand \EOS [0]{\spacefactor3000\relax}%
\providecommand \BibitemShut  [1]{\csname bibitem#1\endcsname}%
\let\auto@bib@innerbib\@empty
\bibitem [{\citenamefont {Birkedal}\ \emph {et~al.}(2004)\citenamefont
  {Birkedal}, \citenamefont {Matchev},\ and\ \citenamefont
  {Perelstein}}]{Birkedal:2004xn}%
  \BibitemOpen
  \bibfield  {author} {\bibinfo {author} {\bibfnamefont {A.}~\bibnamefont
  {Birkedal}}, \bibinfo {author} {\bibfnamefont {K.}~\bibnamefont {Matchev}}, \
  and\ \bibinfo {author} {\bibfnamefont {M.}~\bibnamefont {Perelstein}},\
  }\href {\doibase 10.1103/PhysRevD.70.077701} {\bibfield  {journal} {\bibinfo
  {journal} {Phys.Rev.}\ }\textbf {\bibinfo {volume} {D70}},\ \bibinfo {pages}
  {077701} (\bibinfo {year} {2004})},\ \Eprint
  {http://arxiv.org/abs/hep-ph/0403004} {arXiv:hep-ph/0403004 [hep-ph]}
  \BibitemShut {NoStop}%
\bibitem [{\citenamefont {Bai}\ \emph {et~al.}(2010)\citenamefont {Bai},
  \citenamefont {Fox},\ and\ \citenamefont {Harnik}}]{Bai:2010hh}%
  \BibitemOpen
  \bibfield  {author} {\bibinfo {author} {\bibfnamefont {Y.}~\bibnamefont
  {Bai}}, \bibinfo {author} {\bibfnamefont {P.~J.}\ \bibnamefont {Fox}}, \ and\
  \bibinfo {author} {\bibfnamefont {R.}~\bibnamefont {Harnik}},\ }\href
  {\doibase 10.1007/JHEP12(2010)048} {\bibfield  {journal} {\bibinfo  {journal}
  {JHEP}\ }\textbf {\bibinfo {volume} {1012}},\ \bibinfo {pages} {048}
  (\bibinfo {year} {2010})},\ \Eprint {http://arxiv.org/abs/1005.3797}
  {arXiv:1005.3797 [hep-ph]} \BibitemShut {NoStop}%
\bibitem [{\citenamefont {Goodman}\ \emph {et~al.}(2010)\citenamefont
  {Goodman}, \citenamefont {Ibe}, \citenamefont {Rajaraman}, \citenamefont
  {Shepherd}, \citenamefont {Tait},\ and\ \citenamefont {Yu}}]{Goodman:2010ku}%
  \BibitemOpen
  \bibfield  {author} {\bibinfo {author} {\bibfnamefont {J.}~\bibnamefont
  {Goodman}}, \bibinfo {author} {\bibfnamefont {M.}~\bibnamefont {Ibe}},
  \bibinfo {author} {\bibfnamefont {A.}~\bibnamefont {Rajaraman}}, \bibinfo
  {author} {\bibfnamefont {W.}~\bibnamefont {Shepherd}}, \bibinfo {author}
  {\bibfnamefont {T.~M.~P.}\ \bibnamefont {Tait}}, \ and\ \bibinfo {author}
  {\bibfnamefont {H.-B.}\ \bibnamefont {Yu}},\ }\href {\doibase
  10.1103/PhysRevD.82.116010} {\bibfield  {journal} {\bibinfo  {journal}
  {Phys.Rev.}\ }\textbf {\bibinfo {volume} {D82}},\ \bibinfo {pages} {116010}
  (\bibinfo {year} {2010})},\ \Eprint {http://arxiv.org/abs/1008.1783}
  {arXiv:1008.1783 [hep-ph]} \BibitemShut {NoStop}%
\bibitem [{\citenamefont {An}\ \emph {et~al.}(2015)\citenamefont {An},
  \citenamefont {Echenard}, \citenamefont {Pospelov},\ and\ \citenamefont
  {Zhang}}]{An:2015pva}%
  \BibitemOpen
  \bibfield  {author} {\bibinfo {author} {\bibfnamefont {H.}~\bibnamefont
  {An}}, \bibinfo {author} {\bibfnamefont {B.}~\bibnamefont {Echenard}},
  \bibinfo {author} {\bibfnamefont {M.}~\bibnamefont {Pospelov}}, \ and\
  \bibinfo {author} {\bibfnamefont {Y.}~\bibnamefont {Zhang}},\ }\href@noop {}
  {\  (\bibinfo {year} {2015})},\ \Eprint {http://arxiv.org/abs/1510.05020}
  {arXiv:1510.05020 [hep-ph]} \BibitemShut {NoStop}%
\bibitem [{\citenamefont {An}\ \emph {et~al.}(2013)\citenamefont {An},
  \citenamefont {Huo},\ and\ \citenamefont {Wang}}]{An:2012ue}%
  \BibitemOpen
  \bibfield  {author} {\bibinfo {author} {\bibfnamefont {H.}~\bibnamefont
  {An}}, \bibinfo {author} {\bibfnamefont {R.}~\bibnamefont {Huo}}, \ and\
  \bibinfo {author} {\bibfnamefont {L.-T.}\ \bibnamefont {Wang}},\ }\href
  {\doibase 10.1016/j.dark.2013.03.002} {\bibfield  {journal} {\bibinfo
  {journal} {Phys.Dark Univ.}\ }\textbf {\bibinfo {volume} {2}},\ \bibinfo
  {pages} {50} (\bibinfo {year} {2013})},\ \Eprint
  {http://arxiv.org/abs/1212.2221} {arXiv:1212.2221 [hep-ph]} \BibitemShut
  {NoStop}%
\bibitem [{\citenamefont {Bai}\ and\ \citenamefont
  {Berger}(2013)}]{Bai:2013iqa}%
  \BibitemOpen
  \bibfield  {author} {\bibinfo {author} {\bibfnamefont {Y.}~\bibnamefont
  {Bai}}\ and\ \bibinfo {author} {\bibfnamefont {J.}~\bibnamefont {Berger}},\
  }\href {\doibase 10.1007/JHEP11(2013)171} {\bibfield  {journal} {\bibinfo
  {journal} {JHEP}\ }\textbf {\bibinfo {volume} {1311}},\ \bibinfo {pages}
  {171} (\bibinfo {year} {2013})},\ \Eprint {http://arxiv.org/abs/1308.0612}
  {arXiv:1308.0612 [hep-ph]} \BibitemShut {NoStop}%
\bibitem [{\citenamefont {Liu}\ \emph {et~al.}(2013)\citenamefont {Liu},
  \citenamefont {Shuve}, \citenamefont {Weiner},\ and\ \citenamefont
  {Yavin}}]{Liu:2013gba}%
  \BibitemOpen
  \bibfield  {author} {\bibinfo {author} {\bibfnamefont {J.}~\bibnamefont
  {Liu}}, \bibinfo {author} {\bibfnamefont {B.}~\bibnamefont {Shuve}}, \bibinfo
  {author} {\bibfnamefont {N.}~\bibnamefont {Weiner}}, \ and\ \bibinfo {author}
  {\bibfnamefont {I.}~\bibnamefont {Yavin}},\ }\href {\doibase
  10.1007/JHEP07(2013)144} {\bibfield  {journal} {\bibinfo  {journal} {JHEP}\
  }\textbf {\bibinfo {volume} {1307}},\ \bibinfo {pages} {144} (\bibinfo {year}
  {2013})},\ \Eprint {http://arxiv.org/abs/1303.4404} {arXiv:1303.4404
  [hep-ph]} \BibitemShut {NoStop}%
\bibitem [{\citenamefont {Spergel}\ and\ \citenamefont
  {Steinhardt}(2000)}]{Spergel:1999mh}%
  \BibitemOpen
  \bibfield  {author} {\bibinfo {author} {\bibfnamefont {D.~N.}\ \bibnamefont
  {Spergel}}\ and\ \bibinfo {author} {\bibfnamefont {P.~J.}\ \bibnamefont
  {Steinhardt}},\ }\href {\doibase 10.1103/PhysRevLett.84.3760} {\bibfield
  {journal} {\bibinfo  {journal} {Phys. Rev. Lett.}\ }\textbf {\bibinfo
  {volume} {84}},\ \bibinfo {pages} {3760} (\bibinfo {year} {2000})},\ \Eprint
  {http://arxiv.org/abs/astro-ph/9909386} {arXiv:astro-ph/9909386 [astro-ph]}
  \BibitemShut {NoStop}%
\bibitem [{\citenamefont {{Oh}}\ \emph {et~al.}(2011)\citenamefont {{Oh}},
  \citenamefont {{de Blok}}, \citenamefont {{Brinks}}, \citenamefont
  {{Walter}},\ and\ \citenamefont {{Kennicutt}}}]{DWcore}%
  \BibitemOpen
  \bibfield  {author} {\bibinfo {author} {\bibfnamefont {S.-H.}\ \bibnamefont
  {{Oh}}}, \bibinfo {author} {\bibfnamefont {W.~J.~G.}\ \bibnamefont {{de
  Blok}}}, \bibinfo {author} {\bibfnamefont {E.}~\bibnamefont {{Brinks}}},
  \bibinfo {author} {\bibfnamefont {F.}~\bibnamefont {{Walter}}}, \ and\
  \bibinfo {author} {\bibfnamefont {R.~C.}\ \bibnamefont {{Kennicutt}},
  \bibfnamefont {Jr.}},\ }\href {\doibase 10.1088/0004-6256/141/6/193}
  {\bibfield  {journal} {\bibinfo  {journal} {AJ}\ }\textbf {\bibinfo {volume}
  {141}},\ \bibinfo {eid} {193} (\bibinfo {year} {2011})},\ \Eprint
  {http://arxiv.org/abs/1011.0899} {arXiv:1011.0899} \BibitemShut {NoStop}%
\bibitem [{\citenamefont {{Boylan-Kolchin}}\ \emph {et~al.}(2011)\citenamefont
  {{Boylan-Kolchin}}, \citenamefont {{Bullock}},\ and\ \citenamefont
  {{Kaplinghat}}}]{TBTF1}%
  \BibitemOpen
  \bibfield  {author} {\bibinfo {author} {\bibfnamefont {M.}~\bibnamefont
  {{Boylan-Kolchin}}}, \bibinfo {author} {\bibfnamefont {J.~S.}\ \bibnamefont
  {{Bullock}}}, \ and\ \bibinfo {author} {\bibfnamefont {M.}~\bibnamefont
  {{Kaplinghat}}},\ }\href {\doibase 10.1111/j.1745-3933.2011.01074.x}
  {\bibfield  {journal} {\bibinfo  {journal} {MNRAS}\ }\textbf {\bibinfo
  {volume} {415}},\ \bibinfo {pages} {L40} (\bibinfo {year} {2011})},\ \Eprint
  {http://arxiv.org/abs/1103.0007} {arXiv:1103.0007 [astro-ph.CO]} \BibitemShut
  {NoStop}%
\bibitem [{\citenamefont {Boylan-Kolchin}\ \emph {et~al.}(2011)\citenamefont
  {Boylan-Kolchin}, \citenamefont {Bullock},\ and\ \citenamefont
  {Kaplinghat}}]{TBTF2}%
  \BibitemOpen
  \bibfield  {author} {\bibinfo {author} {\bibfnamefont {M.}~\bibnamefont
  {Boylan-Kolchin}}, \bibinfo {author} {\bibfnamefont {J.~S.}\ \bibnamefont
  {Bullock}}, \ and\ \bibinfo {author} {\bibfnamefont {M.}~\bibnamefont
  {Kaplinghat}},\ }\href {http://arxiv.org/abs/1111.2048} {\  (\bibinfo {year}
  {2011})},\ \Eprint {http://arxiv.org/abs/1111.2048} {1111.2048} \BibitemShut
  {NoStop}%
\bibitem [{\citenamefont {{Walker}}\ and\ \citenamefont
  {{Pe{\~n}arrubia}}(2011)}]{MWslope}%
  \BibitemOpen
  \bibfield  {author} {\bibinfo {author} {\bibfnamefont {M.~G.}\ \bibnamefont
  {{Walker}}}\ and\ \bibinfo {author} {\bibfnamefont {J.}~\bibnamefont
  {{Pe{\~n}arrubia}}},\ }\href {\doibase 10.1088/0004-637X/742/1/20} {\bibfield
   {journal} {\bibinfo  {journal} {\apj}\ }\textbf {\bibinfo {volume} {742}},\
  \bibinfo {eid} {20} (\bibinfo {year} {2011})},\ \Eprint
  {http://arxiv.org/abs/1108.2404} {arXiv:1108.2404} \BibitemShut {NoStop}%
\bibitem [{\citenamefont {{Vogelsberger}}\ \emph {et~al.}(2012)\citenamefont
  {{Vogelsberger}}, \citenamefont {{Zavala}},\ and\ \citenamefont
  {{Loeb}}}]{simulation1}%
  \BibitemOpen
  \bibfield  {author} {\bibinfo {author} {\bibfnamefont {M.}~\bibnamefont
  {{Vogelsberger}}}, \bibinfo {author} {\bibfnamefont {J.}~\bibnamefont
  {{Zavala}}}, \ and\ \bibinfo {author} {\bibfnamefont {A.}~\bibnamefont
  {{Loeb}}},\ }\href {\doibase 10.1111/j.1365-2966.2012.21182.x} {\bibfield
  {journal} {\bibinfo  {journal} {MNRAS}\ }\textbf {\bibinfo {volume} {423}},\
  \bibinfo {pages} {3740} (\bibinfo {year} {2012})},\ \Eprint
  {http://arxiv.org/abs/1201.5892} {arXiv:1201.5892} \BibitemShut {NoStop}%
\bibitem [{\citenamefont {{Rocha}}\ \emph {et~al.}(2013)\citenamefont
  {{Rocha}}, \citenamefont {{Peter}}, \citenamefont {{Bullock}}, \citenamefont
  {{Kaplinghat}}, \citenamefont {{Garrison-Kimmel}}, \citenamefont
  {{O{\~n}orbe}},\ and\ \citenamefont {{Moustakas}}}]{simulation2}%
  \BibitemOpen
  \bibfield  {author} {\bibinfo {author} {\bibfnamefont {M.}~\bibnamefont
  {{Rocha}}}, \bibinfo {author} {\bibfnamefont {A.~H.~G.}\ \bibnamefont
  {{Peter}}}, \bibinfo {author} {\bibfnamefont {J.~S.}\ \bibnamefont
  {{Bullock}}}, \bibinfo {author} {\bibfnamefont {M.}~\bibnamefont
  {{Kaplinghat}}}, \bibinfo {author} {\bibfnamefont {S.}~\bibnamefont
  {{Garrison-Kimmel}}}, \bibinfo {author} {\bibfnamefont {J.}~\bibnamefont
  {{O{\~n}orbe}}}, \ and\ \bibinfo {author} {\bibfnamefont {L.~A.}\
  \bibnamefont {{Moustakas}}},\ }\href {\doibase 10.1093/mnras/sts514}
  {\bibfield  {journal} {\bibinfo  {journal} {MNRAS}\ }\textbf {\bibinfo
  {volume} {430}},\ \bibinfo {pages} {81} (\bibinfo {year} {2013})},\ \Eprint
  {http://arxiv.org/abs/1208.3025} {arXiv:1208.3025} \BibitemShut {NoStop}%
\bibitem [{\citenamefont {{Zavala}}\ \emph {et~al.}(2013)\citenamefont
  {{Zavala}}, \citenamefont {{Vogelsberger}},\ and\ \citenamefont
  {{Walker}}}]{simulation3}%
  \BibitemOpen
  \bibfield  {author} {\bibinfo {author} {\bibfnamefont {J.}~\bibnamefont
  {{Zavala}}}, \bibinfo {author} {\bibfnamefont {M.}~\bibnamefont
  {{Vogelsberger}}}, \ and\ \bibinfo {author} {\bibfnamefont {M.~G.}\
  \bibnamefont {{Walker}}},\ }\href {\doibase 10.1093/mnrasl/sls053} {\bibfield
   {journal} {\bibinfo  {journal} {MNRAS}\ }\textbf {\bibinfo {volume} {431}},\
  \bibinfo {pages} {L20} (\bibinfo {year} {2013})},\ \Eprint
  {http://arxiv.org/abs/1211.6426} {arXiv:1211.6426 [astro-ph.CO]} \BibitemShut
  {NoStop}%
\bibitem [{\citenamefont {Tulin}\ \emph {et~al.}(2013)\citenamefont {Tulin},
  \citenamefont {Yu},\ and\ \citenamefont {Zurek}}]{Tulin:2013teo}%
  \BibitemOpen
  \bibfield  {author} {\bibinfo {author} {\bibfnamefont {S.}~\bibnamefont
  {Tulin}}, \bibinfo {author} {\bibfnamefont {H.-B.}\ \bibnamefont {Yu}}, \
  and\ \bibinfo {author} {\bibfnamefont {K.~M.}\ \bibnamefont {Zurek}},\ }\href
  {\doibase 10.1103/PhysRevD.87.115007} {\bibfield  {journal} {\bibinfo
  {journal} {Phys. Rev.}\ }\textbf {\bibinfo {volume} {D87}},\ \bibinfo {pages}
  {115007} (\bibinfo {year} {2013})},\ \Eprint {http://arxiv.org/abs/1302.3898}
  {arXiv:1302.3898 [hep-ph]} \BibitemShut {NoStop}%
\bibitem [{\citenamefont {Aad}\ \emph {et~al.}(2012)\citenamefont {Aad} \emph
  {et~al.}}]{Aad:2012tfa}%
  \BibitemOpen
  \bibfield  {author} {\bibinfo {author} {\bibfnamefont {G.}~\bibnamefont
  {Aad}} \emph {et~al.} (\bibinfo {collaboration} {ATLAS}),\ }\href {\doibase
  10.1016/j.physletb.2012.08.020} {\bibfield  {journal} {\bibinfo  {journal}
  {Phys. Lett.}\ }\textbf {\bibinfo {volume} {B716}},\ \bibinfo {pages} {1}
  (\bibinfo {year} {2012})},\ \Eprint {http://arxiv.org/abs/1207.7214}
  {arXiv:1207.7214 [hep-ex]} \BibitemShut {NoStop}%
\bibitem [{\citenamefont {Chatrchyan}\ \emph {et~al.}(2012)\citenamefont
  {Chatrchyan} \emph {et~al.}}]{Chatrchyan:2012xdj}%
  \BibitemOpen
  \bibfield  {author} {\bibinfo {author} {\bibfnamefont {S.}~\bibnamefont
  {Chatrchyan}} \emph {et~al.} (\bibinfo {collaboration} {CMS}),\ }\href
  {\doibase 10.1016/j.physletb.2012.08.021} {\bibfield  {journal} {\bibinfo
  {journal} {Phys. Lett.}\ }\textbf {\bibinfo {volume} {B716}},\ \bibinfo
  {pages} {30} (\bibinfo {year} {2012})},\ \Eprint
  {http://arxiv.org/abs/1207.7235} {arXiv:1207.7235 [hep-ex]} \BibitemShut
  {NoStop}%
\bibitem [{\citenamefont {Barbieri}\ \emph {et~al.}(2007)\citenamefont
  {Barbieri}, \citenamefont {Hall}, \citenamefont {Nomura},\ and\ \citenamefont
  {Rychkov}}]{Barbieri:2006bg}%
  \BibitemOpen
  \bibfield  {author} {\bibinfo {author} {\bibfnamefont {R.}~\bibnamefont
  {Barbieri}}, \bibinfo {author} {\bibfnamefont {L.~J.}\ \bibnamefont {Hall}},
  \bibinfo {author} {\bibfnamefont {Y.}~\bibnamefont {Nomura}}, \ and\ \bibinfo
  {author} {\bibfnamefont {V.~S.}\ \bibnamefont {Rychkov}},\ }\href {\doibase
  10.1103/PhysRevD.75.035007} {\bibfield  {journal} {\bibinfo  {journal} {Phys.
  Rev.}\ }\textbf {\bibinfo {volume} {D75}},\ \bibinfo {pages} {035007}
  (\bibinfo {year} {2007})},\ \Eprint {http://arxiv.org/abs/hep-ph/0607332}
  {arXiv:hep-ph/0607332 [hep-ph]} \BibitemShut {NoStop}%
\bibitem [{\citenamefont {Hall}\ \emph {et~al.}(2012)\citenamefont {Hall},
  \citenamefont {Pinner},\ and\ \citenamefont {Ruderman}}]{Hall:2011aa}%
  \BibitemOpen
  \bibfield  {author} {\bibinfo {author} {\bibfnamefont {L.~J.}\ \bibnamefont
  {Hall}}, \bibinfo {author} {\bibfnamefont {D.}~\bibnamefont {Pinner}}, \ and\
  \bibinfo {author} {\bibfnamefont {J.~T.}\ \bibnamefont {Ruderman}},\ }\href
  {\doibase 10.1007/JHEP04(2012)131} {\bibfield  {journal} {\bibinfo  {journal}
  {JHEP}\ }\textbf {\bibinfo {volume} {04}},\ \bibinfo {pages} {131} (\bibinfo
  {year} {2012})},\ \Eprint {http://arxiv.org/abs/1112.2703} {arXiv:1112.2703
  [hep-ph]} \BibitemShut {NoStop}%
\bibitem [{Note1()}]{Note1}%
  \BibitemOpen
  \bibinfo {note} {The name Darkonium in the DM context was first used in \cite
  {Laha:2015yoa} for a stable DM bound state in an asymmetric DM model. Weakly
  coupled DM bound states have been studied in \cite {Pospelov:2008jd,
  MarchRussell:2008tu, Kaplan:2009de,
  Braaten:2013tza,Wise:2014jva}.}\BibitemShut {Stop}%
\bibitem [{\citenamefont {Beltran}\ \emph {et~al.}(2010)\citenamefont
  {Beltran}, \citenamefont {Hooper}, \citenamefont {Kolb}, \citenamefont
  {Krusberg},\ and\ \citenamefont {Tait}}]{Beltran:2010ww}%
  \BibitemOpen
  \bibfield  {author} {\bibinfo {author} {\bibfnamefont {M.}~\bibnamefont
  {Beltran}}, \bibinfo {author} {\bibfnamefont {D.}~\bibnamefont {Hooper}},
  \bibinfo {author} {\bibfnamefont {E.~W.}\ \bibnamefont {Kolb}}, \bibinfo
  {author} {\bibfnamefont {Z.~A.~C.}\ \bibnamefont {Krusberg}}, \ and\ \bibinfo
  {author} {\bibfnamefont {T.~M.~P.}\ \bibnamefont {Tait}},\ }\href {\doibase
  10.1007/JHEP09(2010)037} {\bibfield  {journal} {\bibinfo  {journal} {JHEP}\
  }\textbf {\bibinfo {volume} {09}},\ \bibinfo {pages} {037} (\bibinfo {year}
  {2010})},\ \Eprint {http://arxiv.org/abs/1002.4137} {arXiv:1002.4137
  [hep-ph]} \BibitemShut {NoStop}%
\bibitem [{\citenamefont {Fox}\ \emph {et~al.}(2011)\citenamefont {Fox},
  \citenamefont {Harnik}, \citenamefont {Kopp},\ and\ \citenamefont
  {Tsai}}]{Fox:2011fx}%
  \BibitemOpen
  \bibfield  {author} {\bibinfo {author} {\bibfnamefont {P.~J.}\ \bibnamefont
  {Fox}}, \bibinfo {author} {\bibfnamefont {R.}~\bibnamefont {Harnik}},
  \bibinfo {author} {\bibfnamefont {J.}~\bibnamefont {Kopp}}, \ and\ \bibinfo
  {author} {\bibfnamefont {Y.}~\bibnamefont {Tsai}},\ }\href {\doibase
  10.1103/PhysRevD.84.014028} {\bibfield  {journal} {\bibinfo  {journal} {Phys.
  Rev.}\ }\textbf {\bibinfo {volume} {D84}},\ \bibinfo {pages} {014028}
  (\bibinfo {year} {2011})},\ \Eprint {http://arxiv.org/abs/1103.0240}
  {arXiv:1103.0240 [hep-ph]} \BibitemShut {NoStop}%
\bibitem [{\citenamefont {Fox}\ \emph {et~al.}(2012)\citenamefont {Fox},
  \citenamefont {Harnik}, \citenamefont {Kopp},\ and\ \citenamefont
  {Tsai}}]{Fox:2011pm}%
  \BibitemOpen
  \bibfield  {author} {\bibinfo {author} {\bibfnamefont {P.~J.}\ \bibnamefont
  {Fox}}, \bibinfo {author} {\bibfnamefont {R.}~\bibnamefont {Harnik}},
  \bibinfo {author} {\bibfnamefont {J.}~\bibnamefont {Kopp}}, \ and\ \bibinfo
  {author} {\bibfnamefont {Y.}~\bibnamefont {Tsai}},\ }\href {\doibase
  10.1103/PhysRevD.85.056011} {\bibfield  {journal} {\bibinfo  {journal} {Phys.
  Rev.}\ }\textbf {\bibinfo {volume} {D85}},\ \bibinfo {pages} {056011}
  (\bibinfo {year} {2012})},\ \Eprint {http://arxiv.org/abs/1109.4398}
  {arXiv:1109.4398 [hep-ph]} \BibitemShut {NoStop}%
\bibitem [{\citenamefont {Rajaraman}\ \emph {et~al.}(2011)\citenamefont
  {Rajaraman}, \citenamefont {Shepherd}, \citenamefont {Tait},\ and\
  \citenamefont {Wijangco}}]{Rajaraman:2011wf}%
  \BibitemOpen
  \bibfield  {author} {\bibinfo {author} {\bibfnamefont {A.}~\bibnamefont
  {Rajaraman}}, \bibinfo {author} {\bibfnamefont {W.}~\bibnamefont {Shepherd}},
  \bibinfo {author} {\bibfnamefont {T.~M.~P.}\ \bibnamefont {Tait}}, \ and\
  \bibinfo {author} {\bibfnamefont {A.~M.}\ \bibnamefont {Wijangco}},\ }\href
  {\doibase 10.1103/PhysRevD.84.095013} {\bibfield  {journal} {\bibinfo
  {journal} {Phys. Rev.}\ }\textbf {\bibinfo {volume} {D84}},\ \bibinfo {pages}
  {095013} (\bibinfo {year} {2011})},\ \Eprint {http://arxiv.org/abs/1108.1196}
  {arXiv:1108.1196 [hep-ph]} \BibitemShut {NoStop}%
\bibitem [{\citenamefont {Carpenter}\ \emph {et~al.}(2013)\citenamefont
  {Carpenter}, \citenamefont {Nelson}, \citenamefont {Shimmin}, \citenamefont
  {Tait},\ and\ \citenamefont {Whiteson}}]{Carpenter:2012rg}%
  \BibitemOpen
  \bibfield  {author} {\bibinfo {author} {\bibfnamefont {L.~M.}\ \bibnamefont
  {Carpenter}}, \bibinfo {author} {\bibfnamefont {A.}~\bibnamefont {Nelson}},
  \bibinfo {author} {\bibfnamefont {C.}~\bibnamefont {Shimmin}}, \bibinfo
  {author} {\bibfnamefont {T.~M.~P.}\ \bibnamefont {Tait}}, \ and\ \bibinfo
  {author} {\bibfnamefont {D.}~\bibnamefont {Whiteson}},\ }\href {\doibase
  10.1103/PhysRevD.87.074005} {\bibfield  {journal} {\bibinfo  {journal} {Phys.
  Rev.}\ }\textbf {\bibinfo {volume} {D87}},\ \bibinfo {pages} {074005}
  (\bibinfo {year} {2013})},\ \Eprint {http://arxiv.org/abs/1212.3352}
  {arXiv:1212.3352 [hep-ex]} \BibitemShut {NoStop}%
\bibitem [{\citenamefont {Lin}\ \emph {et~al.}(2013)\citenamefont {Lin},
  \citenamefont {Kolb},\ and\ \citenamefont {Wang}}]{Lin:2013sca}%
  \BibitemOpen
  \bibfield  {author} {\bibinfo {author} {\bibfnamefont {T.}~\bibnamefont
  {Lin}}, \bibinfo {author} {\bibfnamefont {E.~W.}\ \bibnamefont {Kolb}}, \
  and\ \bibinfo {author} {\bibfnamefont {L.-T.}\ \bibnamefont {Wang}},\ }\href
  {\doibase 10.1103/PhysRevD.88.063510} {\bibfield  {journal} {\bibinfo
  {journal} {Phys. Rev.}\ }\textbf {\bibinfo {volume} {D88}},\ \bibinfo {pages}
  {063510} (\bibinfo {year} {2013})},\ \Eprint {http://arxiv.org/abs/1303.6638}
  {arXiv:1303.6638 [hep-ph]} \BibitemShut {NoStop}%
\bibitem [{Note2()}]{Note2}%
  \BibitemOpen
  \bibinfo {note} {Also see \cite {Gupta:2015lfa,Buschmann:2015awa,An:2015pva}
  for another example of producing mediators from the DM
  production.}\BibitemShut {Stop}%
\bibitem [{\citenamefont {Shepherd}\ \emph {et~al.}(2009)\citenamefont
  {Shepherd}, \citenamefont {Tait},\ and\ \citenamefont
  {Zaharijas}}]{Shepherd:2009sa}%
  \BibitemOpen
  \bibfield  {author} {\bibinfo {author} {\bibfnamefont {W.}~\bibnamefont
  {Shepherd}}, \bibinfo {author} {\bibfnamefont {T.~M.~P.}\ \bibnamefont
  {Tait}}, \ and\ \bibinfo {author} {\bibfnamefont {G.}~\bibnamefont
  {Zaharijas}},\ }\href {\doibase 10.1103/PhysRevD.79.055022} {\bibfield
  {journal} {\bibinfo  {journal} {Phys. Rev.}\ }\textbf {\bibinfo {volume}
  {D79}},\ \bibinfo {pages} {055022} (\bibinfo {year} {2009})},\ \Eprint
  {http://arxiv.org/abs/0901.2125} {arXiv:0901.2125 [hep-ph]} \BibitemShut
  {NoStop}%
\bibitem [{\citenamefont {Peskin}\ and\ \citenamefont
  {Schroeder}(1995)}]{Peskin:1995ev}%
  \BibitemOpen
  \bibfield  {author} {\bibinfo {author} {\bibfnamefont {M.~E.}\ \bibnamefont
  {Peskin}}\ and\ \bibinfo {author} {\bibfnamefont {D.~V.}\ \bibnamefont
  {Schroeder}},\ }\href
  {http://www.slac.stanford.edu/spires/find/books/www?cl=QC174.45%3AP4} {\emph
  {\bibinfo {title} {{An Introduction to quantum field theory}}}}\ (\bibinfo
  {year} {1995})\BibitemShut {NoStop}%
\bibitem [{\citenamefont {Kats}\ and\ \citenamefont
  {Schwartz}(2010)}]{Kats:2009bv}%
  \BibitemOpen
  \bibfield  {author} {\bibinfo {author} {\bibfnamefont {Y.}~\bibnamefont
  {Kats}}\ and\ \bibinfo {author} {\bibfnamefont {M.~D.}\ \bibnamefont
  {Schwartz}},\ }\href {\doibase 10.1007/JHEP04(2010)016} {\bibfield  {journal}
  {\bibinfo  {journal} {JHEP}\ }\textbf {\bibinfo {volume} {04}},\ \bibinfo
  {pages} {016} (\bibinfo {year} {2010})},\ \Eprint
  {http://arxiv.org/abs/0912.0526} {arXiv:0912.0526 [hep-ph]} \BibitemShut
  {NoStop}%
\bibitem [{\citenamefont {Kahawala}\ and\ \citenamefont
  {Kats}(2011)}]{Kahawala:2011pc}%
  \BibitemOpen
  \bibfield  {author} {\bibinfo {author} {\bibfnamefont {D.}~\bibnamefont
  {Kahawala}}\ and\ \bibinfo {author} {\bibfnamefont {Y.}~\bibnamefont
  {Kats}},\ }\href {\doibase 10.1007/JHEP09(2011)099} {\bibfield  {journal}
  {\bibinfo  {journal} {JHEP}\ }\textbf {\bibinfo {volume} {09}},\ \bibinfo
  {pages} {099} (\bibinfo {year} {2011})},\ \Eprint
  {http://arxiv.org/abs/1103.3503} {arXiv:1103.3503 [hep-ph]} \BibitemShut
  {NoStop}%
\bibitem [{\citenamefont {Ellwanger}\ \emph {et~al.}(2010)\citenamefont
  {Ellwanger}, \citenamefont {Hugonie},\ and\ \citenamefont
  {Teixeira}}]{Ellwanger:2009dp}%
  \BibitemOpen
  \bibfield  {author} {\bibinfo {author} {\bibfnamefont {U.}~\bibnamefont
  {Ellwanger}}, \bibinfo {author} {\bibfnamefont {C.}~\bibnamefont {Hugonie}},
  \ and\ \bibinfo {author} {\bibfnamefont {A.~M.}\ \bibnamefont {Teixeira}},\
  }\href {\doibase 10.1016/j.physrep.2010.07.001} {\bibfield  {journal}
  {\bibinfo  {journal} {Phys. Rept.}\ }\textbf {\bibinfo {volume} {496}},\
  \bibinfo {pages} {1} (\bibinfo {year} {2010})},\ \Eprint
  {http://arxiv.org/abs/0910.1785} {arXiv:0910.1785 [hep-ph]} \BibitemShut
  {NoStop}%
\bibitem [{\citenamefont {Allanach}\ \emph {et~al.}(2009)\citenamefont
  {Allanach} \emph {et~al.}}]{Allanach:2008qq}%
  \BibitemOpen
  \bibfield  {author} {\bibinfo {author} {\bibfnamefont {B.~C.}\ \bibnamefont
  {Allanach}} \emph {et~al.},\ }\href {\doibase 10.1016/j.cpc.2008.08.004}
  {\bibfield  {journal} {\bibinfo  {journal} {Comput. Phys. Commun.}\ }\textbf
  {\bibinfo {volume} {180}},\ \bibinfo {pages} {8} (\bibinfo {year} {2009})},\
  \Eprint {http://arxiv.org/abs/0801.0045} {arXiv:0801.0045 [hep-ph]}
  \BibitemShut {NoStop}%
\bibitem [{Note3()}]{Note3}%
  \BibitemOpen
  \bibinfo {note} {We thank very helpful discussions with Bibhushan Shakya on
  this point.}\BibitemShut {Stop}%
\bibitem [{\citenamefont {Farina}\ \emph {et~al.}(2014)\citenamefont {Farina},
  \citenamefont {Perelstein},\ and\ \citenamefont {Shakya}}]{Farina:2013fsa}%
  \BibitemOpen
  \bibfield  {author} {\bibinfo {author} {\bibfnamefont {M.}~\bibnamefont
  {Farina}}, \bibinfo {author} {\bibfnamefont {M.}~\bibnamefont {Perelstein}},
  \ and\ \bibinfo {author} {\bibfnamefont {B.}~\bibnamefont {Shakya}},\ }\href
  {\doibase 10.1007/JHEP04(2014)108} {\bibfield  {journal} {\bibinfo  {journal}
  {JHEP}\ }\textbf {\bibinfo {volume} {04}},\ \bibinfo {pages} {108} (\bibinfo
  {year} {2014})},\ \Eprint {http://arxiv.org/abs/1310.0459} {arXiv:1310.0459
  [hep-ph]} \BibitemShut {NoStop}%
\bibitem [{\citenamefont {Franceschini}\ and\ \citenamefont
  {Gori}(2011)}]{Franceschini:2010qz}%
  \BibitemOpen
  \bibfield  {author} {\bibinfo {author} {\bibfnamefont {R.}~\bibnamefont
  {Franceschini}}\ and\ \bibinfo {author} {\bibfnamefont {S.}~\bibnamefont
  {Gori}},\ }\href {\doibase 10.1007/JHEP05(2011)084} {\bibfield  {journal}
  {\bibinfo  {journal} {JHEP}\ }\textbf {\bibinfo {volume} {05}},\ \bibinfo
  {pages} {084} (\bibinfo {year} {2011})},\ \Eprint
  {http://arxiv.org/abs/1005.1070} {arXiv:1005.1070 [hep-ph]} \BibitemShut
  {NoStop}%
\bibitem [{Note4()}]{Note4}%
  \BibitemOpen
  \bibinfo {note} {See \cite {Lu:2013cta} for a similar setup.}\BibitemShut
  {Stop}%
\bibitem [{\citenamefont {Collaboration}(2014)}]{CMS:2014eda}%
  \BibitemOpen
  \bibfield  {author} {\bibinfo {author} {\bibfnamefont {C.}~\bibnamefont
  {Collaboration}} (\bibinfo {collaboration} {CMS}),\ }\href@noop {} {\bibfield
   {journal} {\bibinfo  {journal} {CMS-PAS-HIG-14-013}\ } (\bibinfo {year}
  {2014})}\BibitemShut {NoStop}%
\bibitem [{\citenamefont {Aad}\ \emph {et~al.}(2015{\natexlab{a}})\citenamefont
  {Aad} \emph {et~al.}}]{Aad:2015yza}%
  \BibitemOpen
  \bibfield  {author} {\bibinfo {author} {\bibfnamefont {G.}~\bibnamefont
  {Aad}} \emph {et~al.} (\bibinfo {collaboration} {ATLAS}),\ }\href {\doibase
  10.1140/epjc/s10052-015-3474-x} {\bibfield  {journal} {\bibinfo  {journal}
  {Eur. Phys. J.}\ }\textbf {\bibinfo {volume} {C75}},\ \bibinfo {pages} {263}
  (\bibinfo {year} {2015}{\natexlab{a}})},\ \Eprint
  {http://arxiv.org/abs/1503.08089} {arXiv:1503.08089 [hep-ex]} \BibitemShut
  {NoStop}%
\bibitem [{\citenamefont {Aad}\ \emph {et~al.}(2015{\natexlab{b}})\citenamefont
  {Aad} \emph {et~al.}}]{Aad:2015ufa}%
  \BibitemOpen
  \bibfield  {author} {\bibinfo {author} {\bibfnamefont {G.}~\bibnamefont
  {Aad}} \emph {et~al.} (\bibinfo {collaboration} {ATLAS}),\ }\href {\doibase
  10.1140/epjc/s10052-015-3593-4, 10.1140/epjc/s10052-015-3425-6} {\bibfield
  {journal} {\bibinfo  {journal} {Eur. Phys. J.}\ }\textbf {\bibinfo {volume}
  {C75}},\ \bibinfo {pages} {209} (\bibinfo {year} {2015}{\natexlab{b}})},\
  \bibinfo {note} {[Erratum: Eur. Phys. J.C75,370(2015)]},\ \Eprint
  {http://arxiv.org/abs/1503.04677} {arXiv:1503.04677 [hep-ex]} \BibitemShut
  {NoStop}%
\bibitem [{Note5()}]{Note5}%
  \BibitemOpen
  \bibinfo {note} {Here we assume that the mass splitting between Higgisinos is
  small so that the neutralino and chargino can be treated as stable particles
  before they find each other and annihilate.}\BibitemShut {Stop}%
\bibitem [{\citenamefont {Low}\ and\ \citenamefont {Wang}(2014)}]{Low:2014cba}%
  \BibitemOpen
  \bibfield  {author} {\bibinfo {author} {\bibfnamefont {M.}~\bibnamefont
  {Low}}\ and\ \bibinfo {author} {\bibfnamefont {L.-T.}\ \bibnamefont {Wang}},\
  }\href {\doibase 10.1007/JHEP08(2014)161} {\bibfield  {journal} {\bibinfo
  {journal} {JHEP}\ }\textbf {\bibinfo {volume} {08}},\ \bibinfo {pages} {161}
  (\bibinfo {year} {2014})},\ \Eprint {http://arxiv.org/abs/1404.0682}
  {arXiv:1404.0682 [hep-ph]} \BibitemShut {NoStop}%
\bibitem [{\citenamefont {{Elbert}}\ \emph {et~al.}(2015)\citenamefont
  {{Elbert}}, \citenamefont {{Bullock}}, \citenamefont {{Garrison-Kimmel}},
  \citenamefont {{Rocha}}, \citenamefont {{O{\~n}orbe}},\ and\ \citenamefont
  {{Peter}}}]{coreformat}%
  \BibitemOpen
  \bibfield  {author} {\bibinfo {author} {\bibfnamefont {O.~D.}\ \bibnamefont
  {{Elbert}}}, \bibinfo {author} {\bibfnamefont {J.~S.}\ \bibnamefont
  {{Bullock}}}, \bibinfo {author} {\bibfnamefont {S.}~\bibnamefont
  {{Garrison-Kimmel}}}, \bibinfo {author} {\bibfnamefont {M.}~\bibnamefont
  {{Rocha}}}, \bibinfo {author} {\bibfnamefont {J.}~\bibnamefont
  {{O{\~n}orbe}}}, \ and\ \bibinfo {author} {\bibfnamefont {A.~H.~G.}\
  \bibnamefont {{Peter}}},\ }\href {\doibase 10.1093/mnras/stv1470} {\bibfield
  {journal} {\bibinfo  {journal} {MNRAS}\ }\textbf {\bibinfo {volume} {453}},\
  \bibinfo {pages} {29} (\bibinfo {year} {2015})},\ \Eprint
  {http://arxiv.org/abs/1412.1477} {arXiv:1412.1477} \BibitemShut {NoStop}%
\bibitem [{\citenamefont {{Peter}}\ \emph {et~al.}(2013)\citenamefont
  {{Peter}}, \citenamefont {{Rocha}}, \citenamefont {{Bullock}},\ and\
  \citenamefont {{Kaplinghat}}}]{sidmsimulat}%
  \BibitemOpen
  \bibfield  {author} {\bibinfo {author} {\bibfnamefont {A.~H.~G.}\
  \bibnamefont {{Peter}}}, \bibinfo {author} {\bibfnamefont {M.}~\bibnamefont
  {{Rocha}}}, \bibinfo {author} {\bibfnamefont {J.~S.}\ \bibnamefont
  {{Bullock}}}, \ and\ \bibinfo {author} {\bibfnamefont {M.}~\bibnamefont
  {{Kaplinghat}}},\ }\href {\doibase 10.1093/mnras/sts535} {\bibfield
  {journal} {\bibinfo  {journal} {MNRAS}\ }\textbf {\bibinfo {volume} {430}},\
  \bibinfo {pages} {105} (\bibinfo {year} {2013})},\ \Eprint
  {http://arxiv.org/abs/1208.3026} {arXiv:1208.3026} \BibitemShut {NoStop}%
\bibitem [{\citenamefont {Randall}\ \emph {et~al.}(2008)\citenamefont
  {Randall}, \citenamefont {Markevitch}, \citenamefont {Clowe}, \citenamefont
  {Gonzalez},\ and\ \citenamefont {Bradac}}]{Randall:2007ph}%
  \BibitemOpen
  \bibfield  {author} {\bibinfo {author} {\bibfnamefont {S.~W.}\ \bibnamefont
  {Randall}}, \bibinfo {author} {\bibfnamefont {M.}~\bibnamefont {Markevitch}},
  \bibinfo {author} {\bibfnamefont {D.}~\bibnamefont {Clowe}}, \bibinfo
  {author} {\bibfnamefont {A.~H.}\ \bibnamefont {Gonzalez}}, \ and\ \bibinfo
  {author} {\bibfnamefont {M.}~\bibnamefont {Bradac}},\ }\href {\doibase
  10.1086/587859} {\bibfield  {journal} {\bibinfo  {journal} {Astrophys. J.}\
  }\textbf {\bibinfo {volume} {679}},\ \bibinfo {pages} {1173} (\bibinfo {year}
  {2008})},\ \Eprint {http://arxiv.org/abs/0704.0261} {arXiv:0704.0261
  [astro-ph]} \BibitemShut {NoStop}%
\bibitem [{\citenamefont {Aad}\ \emph {et~al.}(2015{\natexlab{c}})\citenamefont
  {Aad} \emph {et~al.}}]{Aad:2014vea}%
  \BibitemOpen
  \bibfield  {author} {\bibinfo {author} {\bibfnamefont {G.}~\bibnamefont
  {Aad}} \emph {et~al.} (\bibinfo {collaboration} {ATLAS}),\ }\href {\doibase
  10.1140/epjc/s10052-015-3306-z} {\bibfield  {journal} {\bibinfo  {journal}
  {Eur. Phys. J.}\ }\textbf {\bibinfo {volume} {C75}},\ \bibinfo {pages} {92}
  (\bibinfo {year} {2015}{\natexlab{c}})},\ \Eprint
  {http://arxiv.org/abs/1410.4031} {arXiv:1410.4031 [hep-ex]} \BibitemShut
  {NoStop}%
\bibitem [{\citenamefont {Artoni}\ \emph {et~al.}(2013)\citenamefont {Artoni},
  \citenamefont {Lin}, \citenamefont {Penning}, \citenamefont {Sciolla},\ and\
  \citenamefont {Venturini}}]{Artoni:2013zba}%
  \BibitemOpen
  \bibfield  {author} {\bibinfo {author} {\bibfnamefont {G.}~\bibnamefont
  {Artoni}}, \bibinfo {author} {\bibfnamefont {T.}~\bibnamefont {Lin}},
  \bibinfo {author} {\bibfnamefont {B.}~\bibnamefont {Penning}}, \bibinfo
  {author} {\bibfnamefont {G.}~\bibnamefont {Sciolla}}, \ and\ \bibinfo
  {author} {\bibfnamefont {A.}~\bibnamefont {Venturini}},\ }in\ \href
  {http://inspirehep.net/record/1245243/files/arXiv:1307.7834.pdf} {\emph
  {\bibinfo {booktitle} {{Community Summer Study 2013: Snowmass on the
  Mississippi (CSS2013) Minneapolis, MN, USA, July 29-August 6, 2013}}}}\
  (\bibinfo {year} {2013})\ \Eprint {http://arxiv.org/abs/1307.7834}
  {arXiv:1307.7834 [hep-ex]} \BibitemShut {NoStop}%
\bibitem [{\citenamefont {Aad}\ \emph {et~al.}(2015{\natexlab{d}})\citenamefont
  {Aad} \emph {et~al.}}]{Aad:2015zva}%
  \BibitemOpen
  \bibfield  {author} {\bibinfo {author} {\bibfnamefont {G.}~\bibnamefont
  {Aad}} \emph {et~al.} (\bibinfo {collaboration} {ATLAS}),\ }\href {\doibase
  10.1140/epjc/s10052-015-3517-3, 10.1140/epjc/s10052-015-3639-7} {\bibfield
  {journal} {\bibinfo  {journal} {Eur. Phys. J.}\ }\textbf {\bibinfo {volume}
  {C75}},\ \bibinfo {pages} {299} (\bibinfo {year} {2015}{\natexlab{d}})},\
  \bibinfo {note} {[Erratum: Eur. Phys. J.C75,no.9,408(2015)]},\ \Eprint
  {http://arxiv.org/abs/1502.01518} {arXiv:1502.01518 [hep-ex]} \BibitemShut
  {NoStop}%
\bibitem [{\citenamefont {Zhou}\ \emph {et~al.}(2013)\citenamefont {Zhou},
  \citenamefont {Berge}, \citenamefont {Wang}, \citenamefont {Whiteson},\ and\
  \citenamefont {Tait}}]{Zhou:2013raa}%
  \BibitemOpen
  \bibfield  {author} {\bibinfo {author} {\bibfnamefont {N.}~\bibnamefont
  {Zhou}}, \bibinfo {author} {\bibfnamefont {D.}~\bibnamefont {Berge}},
  \bibinfo {author} {\bibfnamefont {L.}~\bibnamefont {Wang}}, \bibinfo {author}
  {\bibfnamefont {D.}~\bibnamefont {Whiteson}}, \ and\ \bibinfo {author}
  {\bibfnamefont {T.}~\bibnamefont {Tait}},\ }\href@noop {} {\  (\bibinfo
  {year} {2013})},\ \Eprint {http://arxiv.org/abs/1307.5327} {arXiv:1307.5327
  [hep-ex]} \BibitemShut {NoStop}%
\bibitem [{Note6()}]{Note6}%
  \BibitemOpen
  \bibinfo {note} {The derivatives in the operator do not cause any suppression
  of decay width, since the typical momentum of decay products is comparable to
  DM mass. One can also consider the scenario where the mediator is a vector
  boson, such as light dark photons. However, the decay in that case suffers
  additional suppressions from the dark photon mass due to Landau-Yang theorem.
  Since the collider study is similar to the scalar channel, we do not consider
  it in this work.}\BibitemShut {Stop}%
\bibitem [{\citenamefont {Curtin}\ \emph {et~al.}(2014)\citenamefont {Curtin},
  \citenamefont {Surujon},\ and\ \citenamefont {Tsai}}]{Curtin:2013qsa}%
  \BibitemOpen
  \bibfield  {author} {\bibinfo {author} {\bibfnamefont {D.}~\bibnamefont
  {Curtin}}, \bibinfo {author} {\bibfnamefont {Z.}~\bibnamefont {Surujon}}, \
  and\ \bibinfo {author} {\bibfnamefont {Y.}~\bibnamefont {Tsai}},\ }\href
  {\doibase 10.1016/j.physletb.2014.10.027} {\bibfield  {journal} {\bibinfo
  {journal} {Phys. Lett.}\ }\textbf {\bibinfo {volume} {B738}},\ \bibinfo
  {pages} {477} (\bibinfo {year} {2014})},\ \Eprint
  {http://arxiv.org/abs/1312.2618} {arXiv:1312.2618 [hep-ph]} \BibitemShut
  {NoStop}%
\bibitem [{\citenamefont {Curtin}\ and\ \citenamefont
  {Tsai}(2014)}]{Curtin:2014afa}%
  \BibitemOpen
  \bibfield  {author} {\bibinfo {author} {\bibfnamefont {D.}~\bibnamefont
  {Curtin}}\ and\ \bibinfo {author} {\bibfnamefont {Y.}~\bibnamefont {Tsai}},\
  }\href {\doibase 10.1007/JHEP11(2014)136} {\bibfield  {journal} {\bibinfo
  {journal} {JHEP}\ }\textbf {\bibinfo {volume} {11}},\ \bibinfo {pages} {136}
  (\bibinfo {year} {2014})},\ \Eprint {http://arxiv.org/abs/1405.1034}
  {arXiv:1405.1034 [hep-ph]} \BibitemShut {NoStop}%
\bibitem [{ATL(2015)}]{ATLAS-CONF-2015-044}%
  \BibitemOpen
  \href {http://cds.cern.ch/record/2052552} {}\bibinfo {type} {Tech. Rep.}\
  \bibinfo {number} {ATLAS-CONF-2015-044}\ (\bibinfo  {institution} {CERN},\
  \bibinfo {address} {Geneva},\ \bibinfo {year} {2015})\BibitemShut {NoStop}%
\bibitem [{\citenamefont {Curtin}\ \emph {et~al.}(2015)\citenamefont {Curtin},
  \citenamefont {Essig}, \citenamefont {Gori},\ and\ \citenamefont
  {Shelton}}]{Curtin:2014cca}%
  \BibitemOpen
  \bibfield  {author} {\bibinfo {author} {\bibfnamefont {D.}~\bibnamefont
  {Curtin}}, \bibinfo {author} {\bibfnamefont {R.}~\bibnamefont {Essig}},
  \bibinfo {author} {\bibfnamefont {S.}~\bibnamefont {Gori}}, \ and\ \bibinfo
  {author} {\bibfnamefont {J.}~\bibnamefont {Shelton}},\ }\href {\doibase
  10.1007/JHEP02(2015)157} {\bibfield  {journal} {\bibinfo  {journal} {JHEP}\
  }\textbf {\bibinfo {volume} {02}},\ \bibinfo {pages} {157} (\bibinfo {year}
  {2015})},\ \Eprint {http://arxiv.org/abs/1412.0018} {arXiv:1412.0018
  [hep-ph]} \BibitemShut {NoStop}%
\bibitem [{\citenamefont {Aad}\ \emph {et~al.}(2014)\citenamefont {Aad} \emph
  {et~al.}}]{Aad:2014yea}%
  \BibitemOpen
  \bibfield  {author} {\bibinfo {author} {\bibfnamefont {G.}~\bibnamefont
  {Aad}} \emph {et~al.} (\bibinfo {collaboration} {ATLAS}),\ }\href {\doibase
  10.1007/JHEP11(2014)088} {\bibfield  {journal} {\bibinfo  {journal} {JHEP}\
  }\textbf {\bibinfo {volume} {11}},\ \bibinfo {pages} {088} (\bibinfo {year}
  {2014})},\ \Eprint {http://arxiv.org/abs/1409.0746} {arXiv:1409.0746
  [hep-ex]} \BibitemShut {NoStop}%
\bibitem [{\citenamefont {Alwall}\ \emph {et~al.}(2011)\citenamefont {Alwall},
  \citenamefont {Herquet}, \citenamefont {Maltoni}, \citenamefont {Mattelaer},\
  and\ \citenamefont {Stelzer}}]{Alwall:2011uj}%
  \BibitemOpen
  \bibfield  {author} {\bibinfo {author} {\bibfnamefont {J.}~\bibnamefont
  {Alwall}}, \bibinfo {author} {\bibfnamefont {M.}~\bibnamefont {Herquet}},
  \bibinfo {author} {\bibfnamefont {F.}~\bibnamefont {Maltoni}}, \bibinfo
  {author} {\bibfnamefont {O.}~\bibnamefont {Mattelaer}}, \ and\ \bibinfo
  {author} {\bibfnamefont {T.}~\bibnamefont {Stelzer}},\ }\href {\doibase
  10.1007/JHEP06(2011)128} {\bibfield  {journal} {\bibinfo  {journal} {JHEP}\
  }\textbf {\bibinfo {volume} {06}},\ \bibinfo {pages} {128} (\bibinfo {year}
  {2011})},\ \Eprint {http://arxiv.org/abs/1106.0522} {arXiv:1106.0522
  [hep-ph]} \BibitemShut {NoStop}%
\bibitem [{\citenamefont {Alloul}\ \emph {et~al.}(2014)\citenamefont {Alloul},
  \citenamefont {Christensen}, \citenamefont {Degrande}, \citenamefont {Duhr},\
  and\ \citenamefont {Fuks}}]{Alloul:2013bka}%
  \BibitemOpen
  \bibfield  {author} {\bibinfo {author} {\bibfnamefont {A.}~\bibnamefont
  {Alloul}}, \bibinfo {author} {\bibfnamefont {N.~D.}\ \bibnamefont
  {Christensen}}, \bibinfo {author} {\bibfnamefont {C.}~\bibnamefont
  {Degrande}}, \bibinfo {author} {\bibfnamefont {C.}~\bibnamefont {Duhr}}, \
  and\ \bibinfo {author} {\bibfnamefont {B.}~\bibnamefont {Fuks}},\ }\href
  {\doibase 10.1016/j.cpc.2014.04.012} {\bibfield  {journal} {\bibinfo
  {journal} {Comput. Phys. Commun.}\ }\textbf {\bibinfo {volume} {185}},\
  \bibinfo {pages} {2250} (\bibinfo {year} {2014})},\ \Eprint
  {http://arxiv.org/abs/1310.1921} {arXiv:1310.1921 [hep-ph]} \BibitemShut
  {NoStop}%
\bibitem [{\citenamefont {Chang}\ \emph {et~al.}(2015)\citenamefont {Chang},
  \citenamefont {Galloway}, \citenamefont {Luty}, \citenamefont {Salvioni},\
  and\ \citenamefont {Tsai}}]{Chang:2014ida}%
  \BibitemOpen
  \bibfield  {author} {\bibinfo {author} {\bibfnamefont {S.}~\bibnamefont
  {Chang}}, \bibinfo {author} {\bibfnamefont {J.}~\bibnamefont {Galloway}},
  \bibinfo {author} {\bibfnamefont {M.}~\bibnamefont {Luty}}, \bibinfo {author}
  {\bibfnamefont {E.}~\bibnamefont {Salvioni}}, \ and\ \bibinfo {author}
  {\bibfnamefont {Y.}~\bibnamefont {Tsai}},\ }\href {\doibase
  10.1007/JHEP03(2015)017} {\bibfield  {journal} {\bibinfo  {journal} {JHEP}\
  }\textbf {\bibinfo {volume} {03}},\ \bibinfo {pages} {017} (\bibinfo {year}
  {2015})},\ \Eprint {http://arxiv.org/abs/1411.6023} {arXiv:1411.6023
  [hep-ph]} \BibitemShut {NoStop}%
\bibitem [{\citenamefont {Laha}(2015)}]{Laha:2015yoa}%
  \BibitemOpen
  \bibfield  {author} {\bibinfo {author} {\bibfnamefont {R.}~\bibnamefont
  {Laha}},\ }\href {\doibase 10.1103/PhysRevD.92.083509} {\bibfield  {journal}
  {\bibinfo  {journal} {Phys. Rev.}\ }\textbf {\bibinfo {volume} {D92}},\
  \bibinfo {pages} {083509} (\bibinfo {year} {2015})},\ \Eprint
  {http://arxiv.org/abs/1505.02772} {arXiv:1505.02772 [hep-ph]} \BibitemShut
  {NoStop}%
\bibitem [{\citenamefont {Pospelov}\ and\ \citenamefont
  {Ritz}(2009)}]{Pospelov:2008jd}%
  \BibitemOpen
  \bibfield  {author} {\bibinfo {author} {\bibfnamefont {M.}~\bibnamefont
  {Pospelov}}\ and\ \bibinfo {author} {\bibfnamefont {A.}~\bibnamefont
  {Ritz}},\ }\href {\doibase 10.1016/j.physletb.2008.12.012} {\bibfield
  {journal} {\bibinfo  {journal} {Phys. Lett.}\ }\textbf {\bibinfo {volume}
  {B671}},\ \bibinfo {pages} {391} (\bibinfo {year} {2009})},\ \Eprint
  {http://arxiv.org/abs/0810.1502} {arXiv:0810.1502 [hep-ph]} \BibitemShut
  {NoStop}%
\bibitem [{\citenamefont {March-Russell}\ and\ \citenamefont
  {West}(2009)}]{MarchRussell:2008tu}%
  \BibitemOpen
  \bibfield  {author} {\bibinfo {author} {\bibfnamefont {J.~D.}\ \bibnamefont
  {March-Russell}}\ and\ \bibinfo {author} {\bibfnamefont {S.~M.}\ \bibnamefont
  {West}},\ }\href {\doibase 10.1016/j.physletb.2009.04.010} {\bibfield
  {journal} {\bibinfo  {journal} {Phys. Lett.}\ }\textbf {\bibinfo {volume}
  {B676}},\ \bibinfo {pages} {133} (\bibinfo {year} {2009})},\ \Eprint
  {http://arxiv.org/abs/0812.0559} {arXiv:0812.0559 [astro-ph]} \BibitemShut
  {NoStop}%
\bibitem [{\citenamefont {Kaplan}\ \emph {et~al.}(2010)\citenamefont {Kaplan},
  \citenamefont {Krnjaic}, \citenamefont {Rehermann},\ and\ \citenamefont
  {Wells}}]{Kaplan:2009de}%
  \BibitemOpen
  \bibfield  {author} {\bibinfo {author} {\bibfnamefont {D.~E.}\ \bibnamefont
  {Kaplan}}, \bibinfo {author} {\bibfnamefont {G.~Z.}\ \bibnamefont {Krnjaic}},
  \bibinfo {author} {\bibfnamefont {K.~R.}\ \bibnamefont {Rehermann}}, \ and\
  \bibinfo {author} {\bibfnamefont {C.~M.}\ \bibnamefont {Wells}},\ }\href
  {\doibase 10.1088/1475-7516/2010/05/021} {\bibfield  {journal} {\bibinfo
  {journal} {JCAP}\ }\textbf {\bibinfo {volume} {1005}},\ \bibinfo {pages}
  {021} (\bibinfo {year} {2010})},\ \Eprint {http://arxiv.org/abs/0909.0753}
  {arXiv:0909.0753 [hep-ph]} \BibitemShut {NoStop}%
\bibitem [{\citenamefont {Braaten}\ and\ \citenamefont
  {Hammer}(2013)}]{Braaten:2013tza}%
  \BibitemOpen
  \bibfield  {author} {\bibinfo {author} {\bibfnamefont {E.}~\bibnamefont
  {Braaten}}\ and\ \bibinfo {author} {\bibfnamefont {H.~W.}\ \bibnamefont
  {Hammer}},\ }\href {\doibase 10.1103/PhysRevD.88.063511} {\bibfield
  {journal} {\bibinfo  {journal} {Phys. Rev.}\ }\textbf {\bibinfo {volume}
  {D88}},\ \bibinfo {pages} {063511} (\bibinfo {year} {2013})},\ \Eprint
  {http://arxiv.org/abs/1303.4682} {arXiv:1303.4682 [hep-ph]} \BibitemShut
  {NoStop}%
\bibitem [{\citenamefont {Wise}\ and\ \citenamefont
  {Zhang}(2014)}]{Wise:2014jva}%
  \BibitemOpen
  \bibfield  {author} {\bibinfo {author} {\bibfnamefont {M.~B.}\ \bibnamefont
  {Wise}}\ and\ \bibinfo {author} {\bibfnamefont {Y.}~\bibnamefont {Zhang}},\
  }\href {\doibase 10.1103/PhysRevD.90.055030, 10.1103/PhysRevD.91.039907}
  {\bibfield  {journal} {\bibinfo  {journal} {Phys. Rev.}\ }\textbf {\bibinfo
  {volume} {D90}},\ \bibinfo {pages} {055030} (\bibinfo {year} {2014})},\
  \bibinfo {note} {[Erratum: Phys. Rev.D91,no.3,039907(2015)]},\ \Eprint
  {http://arxiv.org/abs/1407.4121} {arXiv:1407.4121 [hep-ph]} \BibitemShut
  {NoStop}%
\bibitem [{\citenamefont {Gupta}\ \emph {et~al.}(2015)\citenamefont {Gupta},
  \citenamefont {Primulando},\ and\ \citenamefont {Saraswat}}]{Gupta:2015lfa}%
  \BibitemOpen
  \bibfield  {author} {\bibinfo {author} {\bibfnamefont {A.}~\bibnamefont
  {Gupta}}, \bibinfo {author} {\bibfnamefont {R.}~\bibnamefont {Primulando}}, \
  and\ \bibinfo {author} {\bibfnamefont {P.}~\bibnamefont {Saraswat}},\ }\href
  {\doibase 10.1007/JHEP09(2015)079} {\bibfield  {journal} {\bibinfo  {journal}
  {JHEP}\ }\textbf {\bibinfo {volume} {09}},\ \bibinfo {pages} {079} (\bibinfo
  {year} {2015})},\ \Eprint {http://arxiv.org/abs/1504.01385} {arXiv:1504.01385
  [hep-ph]} \BibitemShut {NoStop}%
\bibitem [{\citenamefont {Buschmann}\ \emph {et~al.}(2015)\citenamefont
  {Buschmann}, \citenamefont {Kopp}, \citenamefont {Liu},\ and\ \citenamefont
  {Machado}}]{Buschmann:2015awa}%
  \BibitemOpen
  \bibfield  {author} {\bibinfo {author} {\bibfnamefont {M.}~\bibnamefont
  {Buschmann}}, \bibinfo {author} {\bibfnamefont {J.}~\bibnamefont {Kopp}},
  \bibinfo {author} {\bibfnamefont {J.}~\bibnamefont {Liu}}, \ and\ \bibinfo
  {author} {\bibfnamefont {P.~A.~N.}\ \bibnamefont {Machado}},\ }\href
  {\doibase 10.1007/JHEP07(2015)045} {\bibfield  {journal} {\bibinfo  {journal}
  {JHEP}\ }\textbf {\bibinfo {volume} {07}},\ \bibinfo {pages} {045} (\bibinfo
  {year} {2015})},\ \Eprint {http://arxiv.org/abs/1505.07459} {arXiv:1505.07459
  [hep-ph]} \BibitemShut {NoStop}%
\bibitem [{\citenamefont {Lu}\ \emph {et~al.}(2014)\citenamefont {Lu},
  \citenamefont {Murayama}, \citenamefont {Ruderman},\ and\ \citenamefont
  {Tobioka}}]{Lu:2013cta}%
  \BibitemOpen
  \bibfield  {author} {\bibinfo {author} {\bibfnamefont {X.}~\bibnamefont
  {Lu}}, \bibinfo {author} {\bibfnamefont {H.}~\bibnamefont {Murayama}},
  \bibinfo {author} {\bibfnamefont {J.~T.}\ \bibnamefont {Ruderman}}, \ and\
  \bibinfo {author} {\bibfnamefont {K.}~\bibnamefont {Tobioka}},\ }\href
  {\doibase 10.1103/PhysRevLett.112.191803} {\bibfield  {journal} {\bibinfo
  {journal} {Phys. Rev. Lett.}\ }\textbf {\bibinfo {volume} {112}},\ \bibinfo
  {pages} {191803} (\bibinfo {year} {2014})},\ \Eprint
  {http://arxiv.org/abs/1308.0792} {arXiv:1308.0792 [hep-ph]} \BibitemShut
  {NoStop}%
\end{thebibliography}%

\end{document}